\documentclass[preprint,12pt]{elsarticle}

\usepackage{amssymb}
\usepackage{microtype}
\usepackage{subfigure}
\usepackage{booktabs} % for professional tables
\usepackage{multirow, rotating} 
\usepackage{tabularx}
\usepackage{longtable}
\usepackage{hyperref}
 \usepackage{xurl}
\hypersetup{breaklinks=true}
\usepackage{amsmath}
\usepackage{mathtools}
\usepackage{amsthm}
\usepackage[capitalize,noabbrev]{cleveref}
\usepackage{pdflscape}
\LTcapwidth=\textwidth

\journal{not yet}

\begin{document}

\begin{frontmatter}



\title{Human Vs. Machines: Who Wins In Semiconductor Market Forecasting?}


\author[TUD,GSofLog,IFX]{Louis Steinmeister\corref{cor1}}
\ead{louis.steinmeister@tu-dortmund.de}
\cortext[cor1]{Corresponding author: Louis Steinmeister}

\author[TUD,GSofLog,RC]{Markus Pauly}

\affiliation[TUD]{organization={Department of Statistics, TU Dortmund University},%Department and Organization
            %addressline={}, 
            city={Dortmund},
            postcode={44227}, 
            state={},
            country={Germany}}
\affiliation[GSofLog]{organization={Graduate School of Logistics},%Department and Organization
            %addressline={}, 
            city={Dortmund},
            postcode={44227}, 
            state={},
            country={Germany}}
\affiliation[IFX]{organization={Infineon Technologies AG},%Department and Organization
            %addressline={}, 
            city={Neubiberg},
            postcode={85579}, 
            state={},
            country={Germany}}
\affiliation[EC]{organization={Research Center Trustworthy Data Science and Security, UA Ruhr},%Department and Organization
            %addressline={}, 
            city={Dortmund},
            postcode={44227}, 
            state={},
            country={Germany}}

\begin{abstract}
“\textit{If you ask ten experts, you will get ten different opinions.}” This common proverb illustrates the common association of expert forecasts with personal bias and lack of consistency. On the other hand, digitization promises consistency and explainability through data-driven forecasts employing machine learning (ML) and statistical models. In the following, we compare such forecasts to expert forecasts from the World Semiconductor Trade Statistics (WSTS), a leading semiconductor market data provider.
\end{abstract}

\begin{graphicalabstract}
\includegraphics[width=\columnwidth]{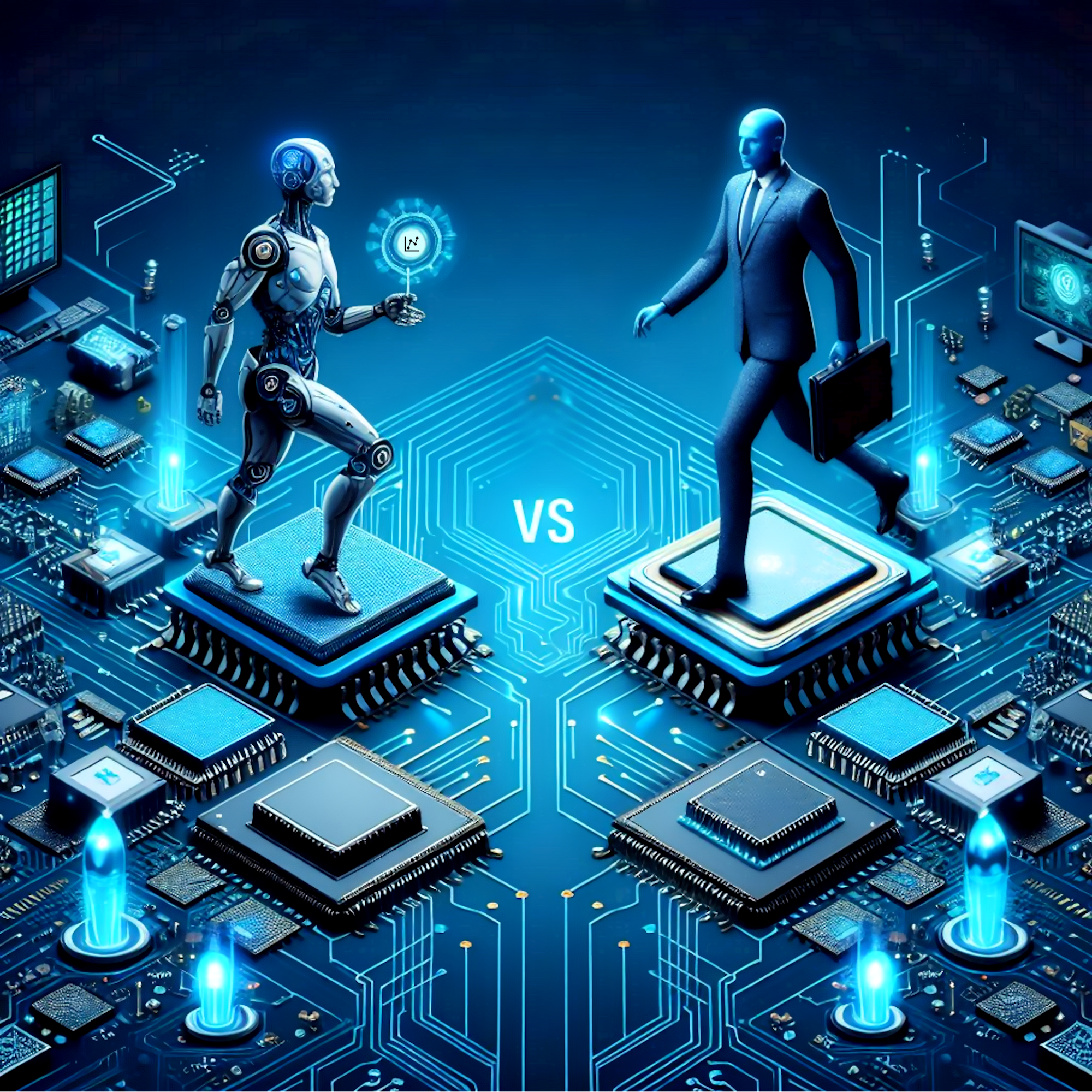}
\end{graphicalabstract}

\begin{highlights}
\item The experts polled by the World Semiconductor Trade Statistics (WSTS), a premier provider of semiconductor market data and forecasts, achieved a high performance in terms of mean error measure and mean ranks compared to data-driven forecasts trained on historic market data up until the official WSTS time stamp.
\item The incorporation of additional available data into the data-driven forecasts resulted in a superior performance of the data-driven models.
\item Contrary to conventional wisdom, the length of the predicted time series did not have an impact on the comparative performance of the data-driven forecasts.
\end{highlights}


\end{frontmatter}


\section{Executive Summary}
\label{ExecutiveSummary}

\textbf{Objectives:} The objective of this paper is to evaluate and contrast expert predictions of the World Semiconductor Trade Statistics (WSTS), a leading semiconductor market data provider, with data-driven forecasts. In this context, we compare expert forecasts with different data-driven forecast approaches with respect to three research hypotheses detailed in the introduction.

\textbf{Motivation:} WSTS plays a crucial role in the semiconductor industry. According to their website, WSTS is the “{\it most respected source of market data and forecasts for the semiconductor industry}” and  their forecasts “{\it are the only ones that leverage the collective experience of the industry’s major players with the market intelligence of a large portion of the semiconductor industry}” \cite{WSTS}. As one of the top providers of comprehensive semiconductor industry data and indicators, WSTS plays a pivotal role in business decision making and industry analyst research. Additionally, the well-being of the semiconductor industry, which lies upstream in the supply chain, has been identified to be a leading indicator for the broader economy \cite{Chow2006}. This highlights the importance of accurate and reliable semiconductor industry forecasts even outside of this specific industry.

\textbf{Methods:} Several popular statistical and ML methods for time series forecasting are evaluated against official forecasts provided by WSTS by means of a time series cross validation.

\textbf{Results:} This paper finds that the expert forecasts provided by WSTS compare favourably to ML forecasts on a quarterly horizon but can nevertheless be enhanced by data driven forecasts. However, the performance of WSTS forecasts is put in perspective when the WSTS algorithmic updates, which are published bi-quarterly, are included. Furthermore, it can be argued that additional information should be incorporated into the forecasts, which results in a clear outperformance of the data-driven methods in comparison to the official WSTS forecasts. This discovery remains consistent regardless of the length of available data points. In other words, the outcome remains unaffected whether we analyze product categories with short histories or those with long histories.

\textbf{Conclusion:}  While WSTS forecasts provide a strong starting point, it is possible to improve the forecast accuracy through data-driven approaches.

\section{Introduction}
\label{sec:Introduction}
In the today's dynamic business landscape, accurate forecasts play an increasingly important role in shaping operative and strategic decisions. The article “Bringing a real-world edge to forecasting” released by McKinsey \& Company in 2020 makes the case that “\textit{[a ‘good’ forecasting process] should be accurate enough to inform a range of critical business decisions – capital reallocation, hiring, strategy, production, and more}” \cite{Agrawal.2020}. In fact, an example of how resource allocation can be improved by increased operational efficiency through more accurate short to mid-term demand forecast is provided by \cite{Pauly2023}.

Furthermore, forecasts play an important role in anticipating technological change \cite{Foster1986, Modis1999SecondLease} and estimating technology and product life cycles, which inform important portfolio and product development decisions \cite{modis1994lifeCycles,Petropoulos2022ForecastingTP,Steinmeister2023}. This is particularly true for the semiconductor industry with long lead times, a dynamic technological environment, and shortening product life cycles \cite{Lv2018, Macher2006, Wu2008}. 

Accurate forecasts of the semiconductor market are relevant to the broader economy.
The global semiconductor market reached sales of \$618 billion in 2022 according to \citet{Alsop.2024b}. This amounted to about 0.61\% of global GDP in 2022 compared to 0.22\% in 1990, highlighting the increasing importance of the industry to the global economy \cite{Alsop.2024, WorldBank.2024}. Semiconductors are ubiquitous in modern life. They enable AI applications, modern defense equipment, data centers, automobiles, wireless communications, all the way to home appliances like your washing machine, gaming console, and electronic toothbrush.
As \citet{Chow2006} observe, the semiconductor market is a leading indicator for the broader economy.

The strategic importance of the semiconductor industry has further been recognized by governments. The US authorized about \$280 billion for research and manufacturing of semiconductors in the US with the CHIPS and Science Act passed in August 2022 \cite{Taylor.2023}. Likewise, the EU subsidizes the industry with roughly 43 billion Euros (roughly \$46.3 billion) \cite{EuropeanCommission.}. Even more impressively, Sam Altman, the current CEO of OpenAI, is seeking \$5 trillion to \$7 trillion of investments to “\textit{boost the world’s chip-building capacity and expand its ability to power AI, among other things}” according to Reuters \cite{Rajan.2024}. To add perspective: this amounts to the combined market capitalization of Microsoft and Apple, the two largest American companies by market capitalization, at the time of writing\footnote{Based on: \url{https://www.tradingview.com/markets/world-stocks/worlds-largest-companies/}. Accessed 03 April 2024.}. 

Industry forecasts that inform important strategic decisions are often expert judgement.
A popular method for consolidating these forecasts is the Delphi method \cite{Linstone.1975Delphi, Armstrong2008, Hyndman.2018Book}. 
However, several researchers have discovered that these collective predictions, also referred to as “wisdom of the crowds" or “collective intelligence," are susceptible to inaccuracies and low precision, particularly when the individuals surveyed are pundits or uninformed laypersons \cite{Modis1999SecondLease, Atanasov.2015}.

However, the industry members polled by WSTS are industry experts and have access to detailed insider information, such as customer orders and sentiment, customer-related project status, the status of customer contracts, and new product development. 
Expert forecasts are generally thought to perform well when systems are complex, dynamic, and available history is sparse \cite{Hyndman.2018Book}. 
The semiconductor industry is heavily intertwined in the global economy, with complex upstream, and downstream supply chains. This means that the semiconductor industry is exposed to the bullwhip effect \cite{lee1997bullwhip}. It is also increasingly subjected to geopolitical considerations. 
While these factors highlight the importance of accurate semiconductor industry forecasts, the dynamic technological environment, geopolitical factors, and the complex supply chains concurrently complicate the data-driven forecast process due to the amount of potentially relevant extrinsic information.
Therefore, expert forecasts from industry insiders queried by WSTS, who possess extensive  access to quantitative and qualitative insider information, are anticipated to be highly reliable. This prompts inquiry into the potential competitiveness of data-driven forecasts, and how or when they could present viable alternatives.
Hence, the following three research hypotheses are to be examined in the following:

\begin{itemize}
    \item[(H1)] Expert forecasts exhibit higher accuracy compared to autoregressive data-driven forecasts.
\end{itemize}

WSTS publishes 
quarterly forecasts for each product category in an alternating pattern: expert forecasts are issued in May and November, while algorithmically computed updates are provided in February and August.
These algorithmic updates are derived from the preceding quarter's results. Industry experts also only have access to official WSTS figures dating back to the previous quarter. Nevertheless, the numbers for the initial month of the forecasted quarter are disclosed simultaneously with the forecasts. Furthermore, it can be anticipated that industry experts possess internal information regarding the first month's data (for a detailed account of the data, see \cref{sec:Data}). This gives rise to the second hypothesis:

\begin{itemize}
    \item[(H2)] The incorporation of additional autoregressive information in data-driven forecasts enhances their competitiveness against expert forecasts.
\end{itemize}

\cref{fig:WSTS_lengths} within \cref{sec:Data} illustrates that certain product categories exhibit significantly shorter historical data compared to others. This observation holds significance as it aligns with the understanding that experts possess the capability to generate accurate forecasts when historical data is limited \cite{Hyndman.2018Book}.  Consequently, this observation motivates the formulation of the third hypothesis:

\begin{itemize}
    \item[(H3)] Expert forecasts outperform data-driven forecasts particularly in the context of short time series. 
\end{itemize}

\textbf{Structure:} 
To investigate these, the
ML and statistical methods used are summarized in the following Section. \cref{sec:ExpSetup} details the experimental setup in terms of the used data (\cref{sec:Data}) and the methodology (\cref{sec:Methodology}). 
The results are discussed in \cref{sec:results}, with \cref{sec:resQuart} addressing (H1), \cref{sec:res2Mo} examining (H2), \cref{sec:resTSLength}   (H3), and finally \cref{sec:Results_additional}, which offers insights of the results at a product category level. \cref{sec:discussion} completes this work with a brief discussion of the findings.

\section{Used Data-Driven Methods}
\label{DDMethods}

This section gives a brief introduction to the data-driven models used in the subsequent analysis. It starts with the description of traditional models based on statistical time series analysis (Section 2.1), continues with ML methods (Section 2.2), and concludes with a brief note on ensemble methods (Section 2.3).

The selection of the models presented here is partially influenced by their performance in the Makridakis Competitions (M-Competitions), which are renowned forecasting competitions conducted on diverse and realistic datasets \cite{Hyndman2020}. 
The results based on the M3 competition are of particular interest: the dataset comprised 3003 individual time series with 14 to 126 observations featuring various levels of seasonality  \cite{Hyndman2020, Makridakis2000}. 

Furthermore, this dataset has served as a benchmark for evaluating popular data-driven forecasting methods. Both 
\citet{Ahmed2010} and \citet{Makridakis2018} 
utilized a subset of the M3 dataset (consisting of 1045 time series) with a minimum length of 81 observations for their analyses.

\subsection{Statistical Models}
\label{StatModels}

Comparisons based on the M3-Competition data have generally favored statistical models over ML approaches \cite{Hyndman2020, Makridakis2018}.
%Statistical models generally performed better in comparisons based on the M3-Competition data \cite{Hyndman2020, Makridakis2018}. 
One suggested reason for this trend is the relatively short length of the time series involved. Unfortunately, this limitation is common in forecasting applications and reflects a realistic constraint. 
The time series examined in this paper, ranging from 92 to 392 monthly observations (further details in \cref{sec:Data}), are longer compared to those studied in the papers based on the M3-Competition, where lengths typically spanned from 81 to 126 observations.  Nevertheless, these lengths are still comparable, especially when contrasted with datasets such as the M5 competition, which feature significantly longer time series, reaching up to approximately 2000 observations \cite{Makridakis2022M5Background}.

\subsubsection{SARIMA}
\label {SARIMA}

The \textbf{s}easonal \textbf{a}uto\textbf{r}egressive \textbf{i}ntegrated \textbf{m}oving \textbf{a}verage (ARIMA) model is a traditional statistical time series model. Its advantages are its interpretability, its wide spread use and that many of its mathematical properties are well known \cite{Brockwell.2002}. 
%In slightly simplified terms
According to \citet{Brockwell.2002}, a time series $X=\{{X}_{t}\}$ is said to be a SARIMA$\left(p,d,q\right)\times\left(P,D,Q\right)$ process with period $s$ if 
$$Y_t := \left(1-B\right)^d\left(1-B^s\right)^D$$
is an causal ARMA process defined as
$$\phi(B)\Phi\left(B^s\right) = \theta(B)\Theta\left(B^s\right),$$
where $B$ is the back-shift operator defined as $BY_t = Y_{-1}$, $\phi(z) = 1- \phi_1z-...-\phi_pz^p$, $\Phi(z) = 1- \Phi_1z-...-\Phi_Pz^P$, $\theta(z)=1+\theta_1z+...+\theta_qz^q$, $\Theta(z)=1+\Theta_1z+...+\Theta_Qz^Q$, and $Z=\left\{Z_{t}\right\}$ being a white noise process.

Generally, an ARMA$\left(p,q\right)$ process 
$Y=\{Y_{t}\}$ is characterized as 
$$Y_t-\sum_{i=1}^{p}{\phi_iY_{t-i}}=Z_t+\sum_{i=1}^{q}{\theta_iZ_{t-i}},$$
with $Z=\left\{Z_{t}\right\}$ being a white noise process. The left-hand side of this equation is the autoregressive part, while the right-hand side is the moving average part (moving average of the error process $Z$). More details on the SARIMA model can be found in \cite{Brockwell.2002}.

These models are often used to describe and to generate data of a wide range of processes. But they can also be used as a predictive model when the parameters are estimated. To this end, we use the \verb+auto.arima+ function of the \verb+forecast+ library in R \cite{Hyndman.2008forecastLibrary}. A similar implementation for Python is available through the \verb+StatsForecast+ library \cite{garza2022statsforecast}.

%There are extensions of the ARIMA model to allow modeling explanatory variables and seasonality, the ARIMAX and SARIMA models. The SARIMAX model combines the benefits of both, ARIMAX and SARIMA. The \verb+auto.arima+, despite its name, searches seasonal models and allows explanatory variables.

The inclusion of the SARIMA model in this work is motivated by its ubiquity in time series analysis and its strong performance on the M3-Competition dataset \cite{Makridakis2018}.

\subsubsection{Simple Exponential Smoothing}
\label{SES}

Exponential smoothing models range back to the 1950’s \cite{Gardner.1985}. Despite their simplicity, they often achieve high predictive performance \cite{Hyndman.2001TimeToMove,Satchell.1995}. Simple exponential smoothing (SES) only requires two quantities: the initial forecast $\hat{X}_\mathbf{0}$ and the smoothing constant $\mathbf{\alpha}$. Consecutive forecasts can then be calculated via
$${\hat{X}}_t=\left(1-\alpha\right){\hat{X}}_{t-1}+\alpha X_{t-1},$$
where $\hat{X}_t$ denotes the one-step forecast for ${X}_{t}$ based upon the history up to ${X}_{t-1}$. An R implementation is available with the \verb+SES+ function of the \verb+forecast+ library \cite{Hyndman.2008forecastLibrary}. A similar implementation for Python is available through the \verb+StatsForecast+ library \cite{garza2022statsforecast}.

SES' simplicity, ease of implementation, and computational efficiency make it a popular forecasting tool for practitioners. Additionally, the model performed well on the M3 and M5 Competition datasets \cite{Makridakis2018, Makridakis2022}.

\subsubsection{Error, Trend, and Seasonality}
\label{ETS}

Error, Trend, and Seasonality (ETS) approaches are a flexible class of exponential smoothing models that go beyond SES (see above). As their name suggests, they are capable of modelling time series with trends and seasonality \cite{Hyndman.2018Book}. ETS was the best performing model in the
\citet{Makridakis2018}
comparison based on the M3-Competition data.  It is also implemented as part of the \verb+forecast+ library in R \cite{Hyndman.2008forecastLibrary}. As for the previous two models, a similar implementation for Python is available through the \verb+StatsForecast+ library \cite{garza2022statsforecast}.

\subsection{ML Models}
\label{MLModels}

This subsection introduces the used ML models. 
\citet{Makridakis2018} found that ML methods performed worse than classical statistical models for relatively short time series  -  a finding that was confirmed by \citet{Cerqueira2022}. 
This is particularly the case for artificial neural networks and deep learning models, which are well known to require large sample sizes to produce the desired results \cite{goodfellow2016deep}. This was also verified by the NN3-Competition, which extended the M3-Competition to include neural network approaches \cite{Hyndman2020, crone2011advances}. Therefore, following \cite{Cerqueira2022}, this paper does not discuss neural network models despite their considerable popularity in recent years. Likewise, boosting models are not included.

\subsubsection{Random Forest}
\label{RF}

Random forests (RF) are a bagging algorithm, a specific kind of ensemble learning, which combines the outputs of multiple decision trees \cite{Breiman.2001}. Ahmed et al. and Makridakis et al. included Categorization and Regresssion Trees (CART), which generate single decision trees for regression or classification purposes \cite{breiman1984cart}. However, since its introduction, RF has proven to be an incredibly versatile and successful model for both regression and classification \cite{Biau.2016,huang2020travel,grinsztajn2022why}. We use the \verb+ranger+ implementation of this model as provided by \cite{Wright.2017}. An alternative Python implementation is available through the \verb+skranger+ library. During each CV-step (see below), a grid search was conducted to tune the three hyper-parameters \cite{Probst2019}:
\begin{align*}
    mtry & \in \{2,7,12,16,23\} \\
    min.node.size & \in \{5,7,10\} \\
    splitrule & \in \{variance, extratrees\}.
\end{align*}

\subsubsection{Extremely Randomized Trees}
\label{extraTree}
Extremely Randomized Trees (ExtraTrees, also referred to as ET for brevity) is a model similar to Random Forest (see above). The difference lies in randomizing the splitting point and the feature to split on during training. It is computationally more efficient and promises greater accuracy on a range of problems \cite{Geurts.2006}. The \verb+ranger+ library is also used for this model \cite{Wright.2017}. However, in contrast to the RF model, the parameter \verb|splitrule| remained fixed as \verb|extratrees|. A Python implementation is available through \verb|sklearn| \cite{scikit-learn}. During each CV-step, a grid search was conducted to tune the hyper-parameters \cite{Probst2019}:
\begin{align*}
    mtry & \in \{2,7,12,16,23\} \\
    min.node.size & \in \{5,7,10\}. 
\end{align*}

\subsubsection{Gaussian Processes Regression}
\label{GPR}
Gaussian Processes Regression (GPR) is a probabilistic regression model incorporating Bayesian ideas: A prior distribution of possible regression functions is narrowed down as evidence (observed data points) are incorporated to yield a posterior distribution \cite{Wang.2023}. It has been shown that GPR can be viewed as a limit of many artificial neural network designs and ARMA processes can be viewed as a Gaussian process under the right conditions \cite{Williams1995GPR}. Furthermore, due to their probabilistic nature, GPR easily provides uncertainty quantification  for the forecasts in terms of
prediction intervals. 
GPR showed a promising performance on the M3-Competition dataset \cite{Ahmed2010, Makridakis2018}.
This analysis uses the GPR implementation of the \verb+kernlab+ R library \cite{Karatzoglou.2004}. The model is also implemented for Python in the \verb|sklearn| library \cite{scikit-learn}.

\subsubsection{K-Nearest Neighbors}
\label{KNN}
K-Nearest Neighbors (KNN) is a popular non-parametric model classification and regression model. It bases estimates on the K nearest neighbors in the covariate space \cite{Cover.1967, Fix.1989}. K represents a hyper-parameter to be tuned. For regression, a mean of these K nearest neighbors is usually used as the predictor in regression. This model can be employed with a kernel. In the following, the kernel is automatically chosen.
While KNN has not been among the best performing models on the M3-Competition data \cite{Ahmed2010, Makridakis2018}, it is nevertheless popular as simple a non-parametric model.
Here, the implementation of the \verb+kknn+ library is used \cite{kknn}. KNN is implemented for Python as the \verb|NearestNeighbors| model in the \verb|sklearn| library \cite{scikit-learn}. During each CV-step, a grid search was conducted to tune hyper-parameters among:
\begin{align*}
    K & \in \{1,2,3,4,5,7,9\} \\
    distance & \in \{L1,L2,L3\},
\end{align*}
where the $L1, L2, L3$ distances are given by
\begin{align*}
    d_{L^p}(x,y) := ||x-y||_p
\end{align*}
given that $||x||_p := \left(\sum_{i=1}x_i^p \right) ^{1/p}$
is the $\mathcal{L}^p$ norm.

\subsubsection{Support Vector Regression}
\label{SVR}

Support Vector Regression (SVR) was proposed as an extension to the classical Support Vector Machine (SVM) for classification \cite{SVM1995} to tackle regression problems \cite{svr1996}. Instead minimizing all residuals, such as in ordinary least square regression, the distance of observations outside a margin of error to this margin of error (the $\epsilon$-insensitive tube) are minimized. Analogous to SVMs, these observations are called support vectors, because the regressor only depends on these observations. To model non-linear dependencies, kernels can be used \cite{EfficientLearningMachines}. We use the radial kernel to add another non-linear method. The used model is implemented in the \verb|kernlab| library for R or the \verb|sklearn| library for Python \cite{Karatzoglou.2004, scikit-learn}. A grid search was conducted to tune hyper-parameters among:
\begin{align*}
    \sigma & \in \left\{\frac{1}{16}, \frac{1}{8}, \frac{1}{4}, \frac{1}{2}, 1 \right\} \\
    C & \in \left\{\frac{1}{4}, \frac{1}{2}, 1, 2, 4 \right\}.
\end{align*}

\subsection{Ensemble}
\label{Ensemble}
Ensemble models consist of several individual models which are combined to produce a single output \cite{Opitz.1999}. In addition to the tree-based ensembles Random Forests and ExtraTrees, this paper also analyses a simple ensemble of all the employed data-driven models by taking the median prediction.

\section{Experimental Setup}
\label{sec:ExpSetup}

This section provides an overview over the data used in the analysis (\cref{sec:Data}) and the methodology behind the time series training and model evaluation (\cref{sec:Methodology}).

\subsection{Data}
\label{sec:Data}

This paper analyzes time series aggregated sales of 110 WSTS product categorizations, which were reported monthly and measured in USD. The data is accessible through a WSTS membership \footnote{https://www.wsts.org/61/membership} or a subscription \footnote{https://www.wsts.org/61/subscription}. A major challenge was the consistency of the data: 
\begin{figure}[ht]
%\vskip 0.2in
\begin{center}
\centerline{\includegraphics[width=\columnwidth]{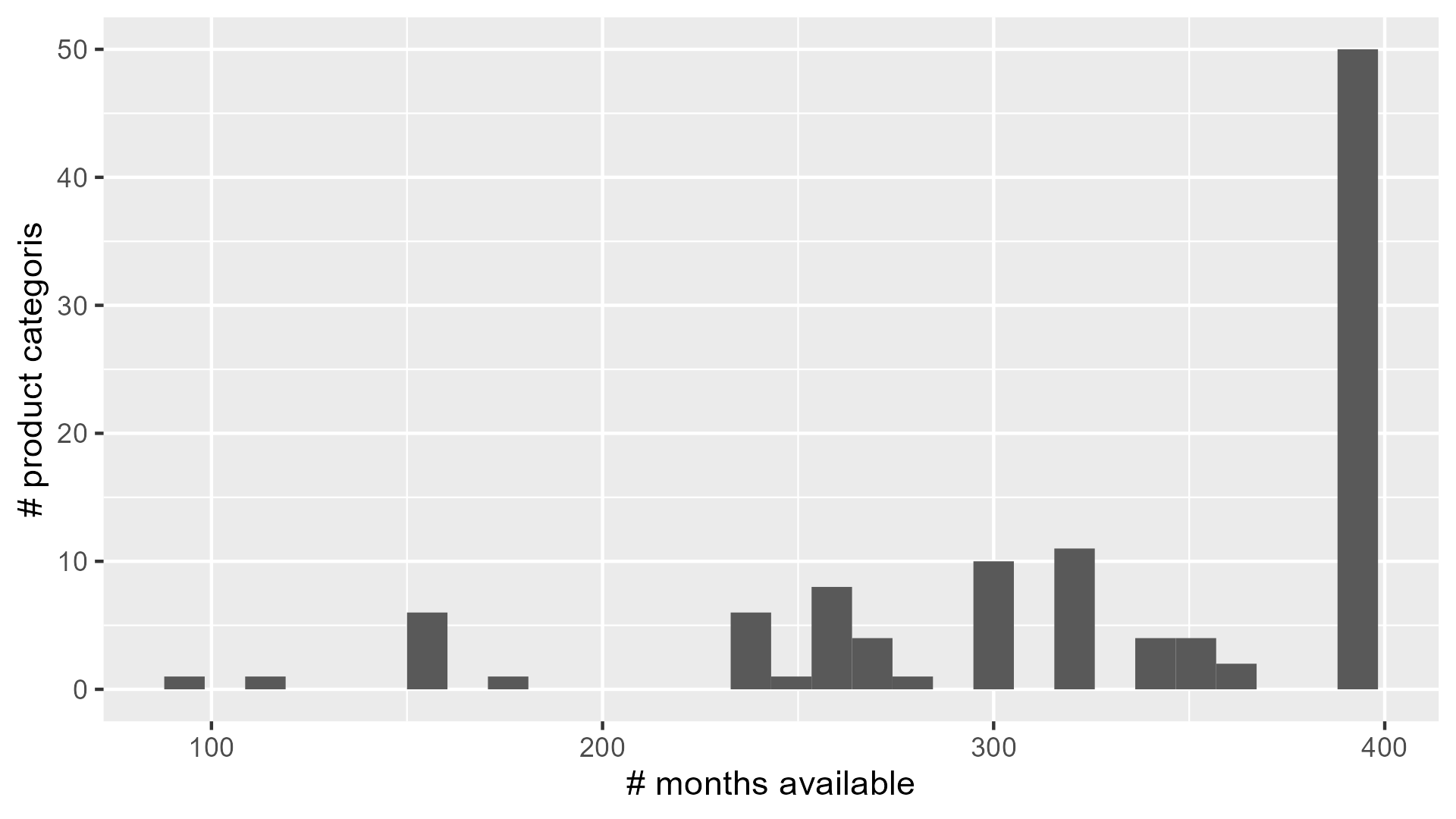}}
\caption{Histogram of time series lengths.}
\label{fig:WSTS_lengths}
\end{center}
%\vskip -0.2in
\end{figure}

Given the dynamic nature of the semiconductor market, product categorizations changed over time. The historical consistency of the current product categorizations was investigated and resolved dating back to Jan 2010 by merging the C7a and C7b product categorizations to C7 (Field Effect General Purpose Power Transistors), the P51 and P52 categorizations into P5 (Automotive and General Purpose MCU), and the L7a, L7b, L7c, L7d and L7f categorizations into L7a/b/c/d/f (Wireless Communication Total). The newest category (F10) dates back to Jan 2016. ~\cref{fig:WSTS_lengths} provides an overview of the months of history of the categorizations as used in the analysis. These are the categories which are consistent until August 2023 and they provide a comprehensive overview of the semiconductor market. We refer to \citet{WSTS} for an exact description of the market and each category. 

\begin{landscape}
\begin{figure*}[ht]
\vskip 0.2in
\centering
\includegraphics[width=\linewidth]{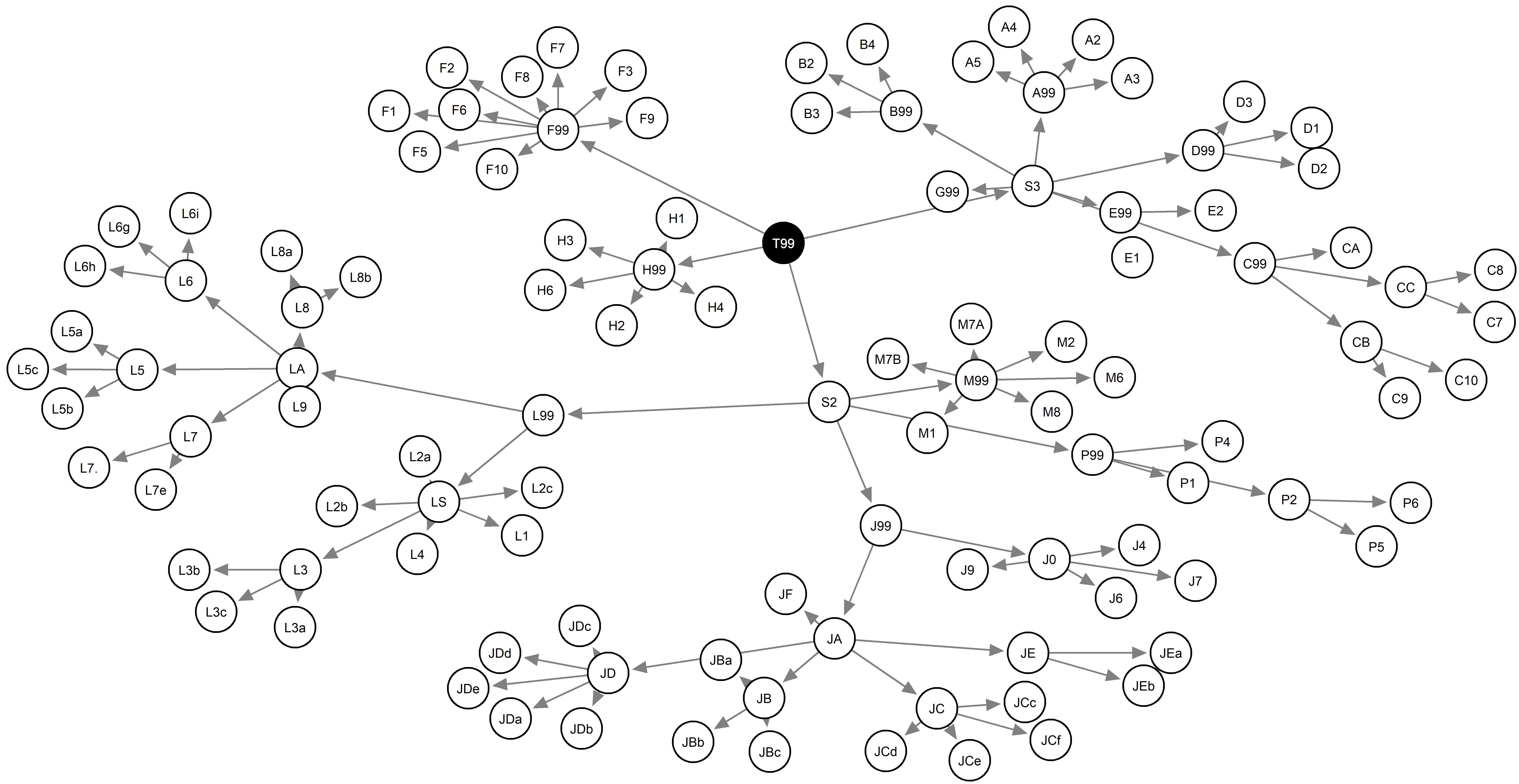}
\caption{WSTS product categorization hierarchy. The highest aggregation level is T99, the node colored in black with white print, slightly to the top of the center of the illustration. The arrows point to the subsumed product categories.}
\label{fig:WSTS_Hierarhy}
\vskip -0.2in
\end{figure*}
\end{landscape}

Generally, the categories positioned higher in the hierarchy exhibit greater consistency. ~\cref{fig:WSTS_Hierarhy} illustrates the hierarchical structure of the product categorizations (starting with T99 as the highest aggregation). To effectively incorporate seasonal components, model fitting necessitated a minimum of 24 months (two seasonal cycles worth) of training data. The shortest time series comprised 92 monthly data points, thus leaving 68 months (or about 17.4\% of the complete time series from 1991) as a test set for the first step of the rolling time series cross-validation (CV, see below). Hence, the first training set spanned all data from January 1991 (or whenever available) to December 2017. Consequently, the test set spanned the time frame from January 2018 to August 2023. 

Official forecasts from WSTS were released quarterly from Q1 2016 to Q3 2023 (midway through the first forecasted quarter). Expert forecasts are consolidated during a global WSTS meeting twice a year – each May and November. 
WSTS additionally issues forecast updates semiannually, in February and August, derived from the preceding meeting's expert forecasts and updated algorithmically with new data reported for the prior quarter.
For example, the forecast for Q2 2024 relies on the upcoming global WSTS meeting scheduled for May 20th-23rd, while the February 2024 forecast update drew upon forecasts from the November 2023 meeting and the reported data from Q4 2023. Thus, 11 expert forecasts and 12 updated forecasts were considered in the analyzed time span from January 2018 to August 2023.

\subsection{Methodology}
\label{sec:Methodology}
{\bf Training and evaluating the ML models:} As described in the last subsection, each time series is split into first training and test sections. Time series with a longer available history consequently have more data points for training than shorter time series, i.e. newer product categories. 

\begin{figure}[ht]
\begin{center}
\centerline{\includegraphics[width=\columnwidth]{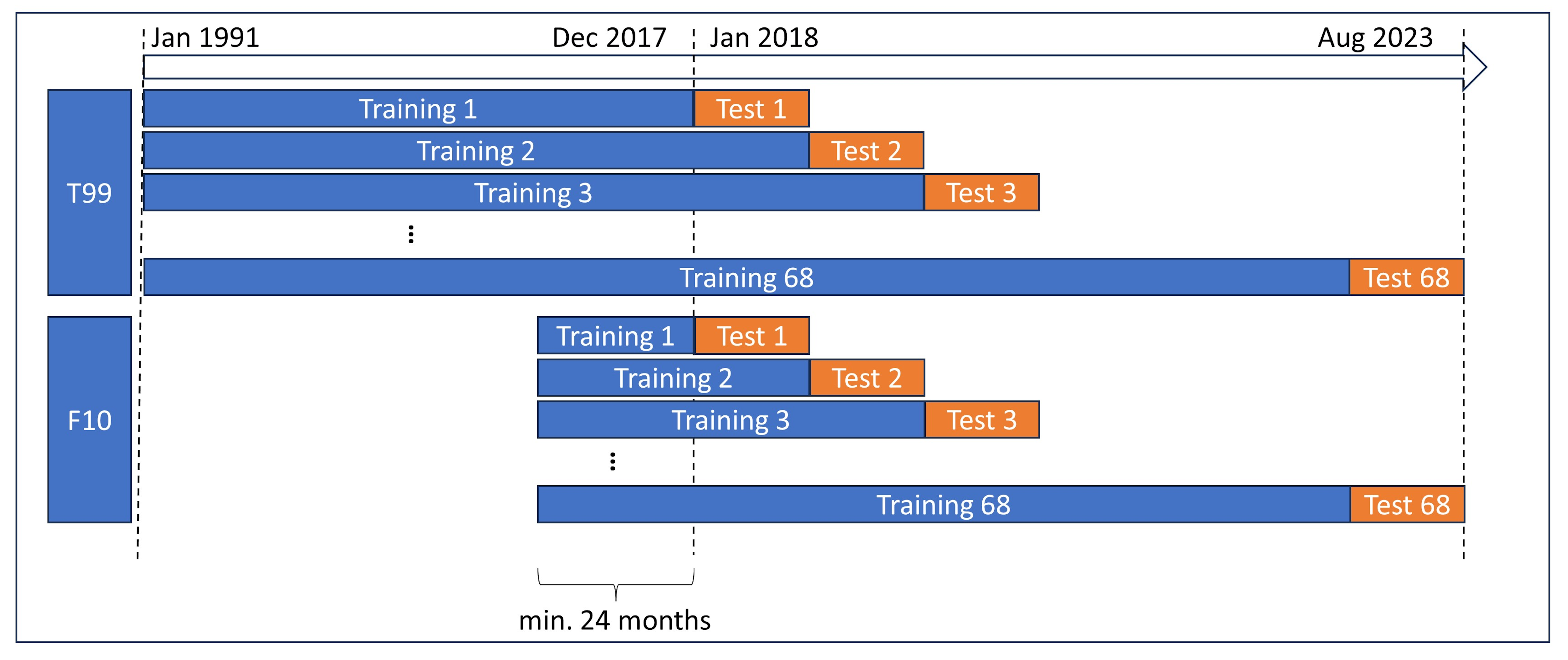}}
\caption{Illustration of time series cross validation for two product categories with differing lengths: T99 has a much longer history of reported aggregate sales than F10.}
\label{fig:CV}
\end{center}
\end{figure}

To obtain reliable forecast performance estimates for all of them, rolling time series cross-validation (CV) as in \citet{Hyndman.2018Book} is performed on each time series and for each model. This is illustrated in \cref{fig:CV}, which also shows the most extreme training periods for the different time series (only 24 months for the first forecast of category F10 up to 390 months for the last forecast of T99).

Within each iteration, the training data is automatically transformed with the Box-Cox transformation \cite{Box.1964} given by
$$
X_t^{(\lambda)} = \begin{cases}
 (X_t^\lambda-1)/\lambda & \lambda \neq 0,\\
 \log(X_t), & \lambda=0.
\end{cases}
$$

This transformation is incorporated and automatically estimated by the used libraries “forecast” and “caretForecast” \cite{Hyndman.2008forecastLibrary, Akay.2022}. Applying the Box-Cox transformation is standard practise, especially when residual distributions are skewed, and when non-negative forecasts are desired \cite{Hyndman.2018Book}. The considered time series report aggregated sales in USD. Thus, there is unlimited upside potential whereas the lower bound is always zero since all time series are positive. Note that $\lambda$ can always be chosen close to one if the transformation is not particularly helpful. Automatically applying it to all cases therefore doesn’t hurt. This is also the default setting in the \verb+forecast+ library \cite{Hyndman.2008forecastLibrary}.

Additionally, hyper-parameters of all data-driven models introduced in Section~\ref{DDMethods} are optimized using the default grid search setting of the “caret” and “caretForecast” libraries \cite{Kuhn.2008, Akay.2022} if no hyper-parameter optimization is conducted through the learning algorithm (one example where hyper-parameters are tuned automatically is the SARIMA model using the “auto.arima” function).
After each training iteration, a forecast is generated up to three months in advance. These forecasts can be compared against the reported numbers, providing an estimate of the performance of the model, and the WSTS forecasts. 

{\bf Evaluation and comparison with WSTS’ forecasts: } The forecasts provided by WSTS are evaluated by type (``meeting" corresponding to the two WSTS expert forecast per year, ``alg. update" corresponding to the two WSTS algorithmic forecasts per year and ``overall" for all four WSTS forecasts per year) and compared to the corresponding data-driven forecasts. The first comparison evaluates all forecasts with a forecasting horizon of $h=3$ months. This corresponds to the information timestamp available to WSTS’ updating algorithm and the industry experts. 

At the same time, it is safe to assume that the industry experts have access to the sales data (among many more) of the first month of each quarter when the meeting convenes (the meeting is held in about the middle of the first quarter to be forecasted). Additionally, the forecasts are disclosed at the same time as the first month’s results are published. Hence, the information time stamp of the forecast and the first month’s results is the same. Utilizing this information reduces the required forecast horizon to $h=2$ months. This is analyzed in a second step to investigate whether using more of the available information might boost the forecasting accuracy.

Similar to \cite{Hyndman2006anotherlook,Pauly2023}, the forecasting accuracy is evaluated using the mean squared error (MSE), mean absolute percentage error (MAPE), and the mean absolute error (MAE). These are given by the following equations
\begin{align*}
  MSE &= \frac{1}{T}\sum_{t=1}^T (\hat{X}_t - X_t)^2  \\
  MAPE &= \frac{1}{T}\sum_{t=1}^T |\hat{X}_t - X_t|/X_t  \\
  MAE &= \frac{1}{T}\sum_{t=1}^T |\hat{X}_t - X_t|,
\end{align*}
where again $\hat{X}_t$ is the one-step forecast for the $t$-th observation ${X}_{t}$ of the test set and $T=68$ is the number of evaluated forecasts. We note that the MAPE is applicable as the values of all time series are positive.

\section{Results}
\label{sec:results}
As discussed in \cref{sec:Methodology}, first, the quarterly forecasts ($h = 3$ months) are discussed in \cref{sec:resQuart} to examine the first research hypothesis (H1) that expert forecasts exhibit higher accuracy compared to data-driven forecasts. This is followed by a comparison with the model performance when an additional month of available information ($h = 2$ months) is incorporated in \cref{sec:res2Mo}, addressing  the second research hypothesis (H2). Lastly, the results are contrasted by time series length

\subsection{Quarterly Forecast Performance}
\label{sec:resQuart}

WSTS' forecasts are provided on a quarterly basis. Each quarter, the previous quarter's numbers are known when the forecasts are compiled. Therefore, as a first step, the data-driven models are compared against WSTS' forecasts on a forecasting horizon of $h = 3$ months, i.e. one quarter. 

\begin{landscape}
\begin{table*}[htbp]
  \centering
  \caption{Average performance of the data-driven forecasts across all 110 product categories and relative to the World Semiconductor Trade Statistics' (WSTS). Each row refers to a different error measure, sorted by WSTS' forecast type: algorithmic update, meeting (expert forecast), and overall. Lower values are preferable. The best value per row is bold and italic.}
    \begin{tabular}{|cccccccccccc|}
    \multicolumn{1}{r}{} &       & \textbf{WSTS} & \textbf {SARIMA} & \textbf{ETS} & \textbf{ET} & \textbf{GPR} & \textbf{KNN} & \textbf{RF} & \textbf{SES} & \textbf{SVM} & \multicolumn{1}{c}{\textbf{Ensemble}} \\
    \hline
    \multicolumn{1}{|c}{\multirow{3}[2]{*}{\textbf{Alg. \newline{}Update}}} & MSE   &              1.00  & \textit{\textbf{            0.34 }} &              0.55  &              1.55  &              0.37  &              4.47  &                 1.49  &              0.94  &              3.20  &          0.64  \\
          & MAPE  & \textit{\textbf{            1.00 }} &              1.01  &              1.01  &              1.20  &              1.01  &              1.71  &                 1.18  &              1.06  &              1.69  &          1.04  \\
          & MAE   &              1.00  & \textit{\textbf{            0.75 }} &              0.81  &              1.34  &              0.77  &              2.14  &                 1.29  &              1.08  &              1.94  &          0.95  \\
    \hline
    \multirow{3}[2]{*}{\textbf{Meeting}} & MSE   & \textit{\textbf{            1.00 }} &              1.61  &              1.57  &              2.78  &              1.33  &              7.23  &                 2.49  &              1.79  &              6.91  &          1.77  \\
          & MAPE  & \textit{\textbf{            1.00 }} &              1.14  &              1.09  &              1.32  &              1.13  &              1.76  &                 1.28  &              1.10  &              1.82  &          1.14  \\
          & MAE   & \textit{\textbf{            1.00 }} &              1.22  &              1.22  &              1.53  &              1.17  &              2.32  &                 1.47  &              1.24  &              2.32  &          1.28  \\
    \hline
    \multirow{3}[2]{*}{\textbf{Overall}} & MSE   &              1.00  &              0.73  &              0.86  &              1.93  & \textit{\textbf{            0.66 }} &              5.31  &                 1.79  &              1.20  &              4.33  &          0.98  \\
          & MAPE  & \textit{\textbf{            1.00 }} &              1.08  &              1.05  &              1.26  &              1.07  &              1.73  &                 1.23  &              1.08  &              1.75  &          1.09  \\
          & MAE   &              1.00  &              0.97  &              1.00  &              1.42  & \textit{\textbf{            0.95 }} &              2.22  &                 1.37  &              1.15  &              2.11  &          1.10  \\
    \hline
    \end{tabular}%
  \label{tab:3moErr}%
\end{table*}% 
\end{landscape}

\cref{tab:3moErr} presents the average performance of data-driven forecasts across 110 product categories, relative to the average performance of forecasts provided by the World Semiconductor Trade Statistics (WSTS). Hence, the first data column (for WSTS) always reads 1. Each row represents a different error measure, organized according to WSTS' forecast types: algorithmic update, meeting (expert forecast), and overall. Lower values are preferable in all cases.

For WSTS' ``Algorithmic Update" forecasts, three error measures are reported: Mean Squared Error (MSE), Mean Absolute Percentage Error (MAPE), and Mean Absolute Error (MAE) -- see \cref{sec:Methodology}. Among these measures, the data-driven forecasts outperform WSTS in terms of MSE and MAE, with the best-performing model indicated by bold and italic formatting. The SARIMA model exhibits the lowest MSE (0.34 relative to WSTS) and MAE (0.75), suggesting superior performance in this category. Almost as good was the GPR model (relative MSE of 0.37 and MAE of 0.77). In term of MAPE, WSTS performed slightly better than several data driven forecast methods: SARIMA, ETS, and GPR each scored 1\% worse. Overall, these results indicate potential for improvement of WSTS' algorithmic update protocol.

Similarly, for the ``Meeting" forecasts, the same three error measures are provided. 
Consistent with the first research hypothesis (H1), industry experts demonstrated superior performance across all three error measures.  
However, among the data-driven models, the best performers were GPR, exhibiting a 33\% increase in MSE compared to expert forecasts, while ETS and SES displayed 9\% and 10\% higher MAPE values respectively. Additionally, GPR, again, was the best performing data-driven model in terms of MAE, showing a 17\% higher MAE relative to the expert forecasts.

\cref{tab:3moErr} concludes with the ``Overall" forecasts, providing the average performance of all models relative to the average performance of 
the combined algorithmic and expert forecasts by WSTS (from 2 meeting and 2 algorithmic forecast per year).
Once more, the GPR model emerges as the 
top-performing data-driven model, showing a 34\% improvement in MSE and a 5\% enhancement in MAE compared to WSTS.
However, WSTS outperforms all models in terms of MAPE, with ETS, the top-performing data-driven model, recording a 5\% higher error than WSTS.

% 3 mo ranks
\begin{landscape}
\begin{table*}[htbp]
  \centering
  \caption{Mean ranks of the forecasts with horizon h = 3 months across all 110 product categories. Each row refers to a different error measure, sorted by WSTS' forecast type: algorithmic update, meeting (expert forecast), and overall. Lower values are preferable. The best value per row is bold and italic.}
    \begin{tabular}{|cccccccccccc|}
    \multicolumn{1}{r}{} &       & \textbf{WSTS} & \textbf{ARIMA} & \textbf{ETS} & \textbf{ET} & \textbf{GPR} & \textbf{KNN} & \textbf{RF} & \textbf{SES} & \textbf{SVM} & \multicolumn{1}{c}{\textbf{Ensemble}} \\
    \hline
    \multicolumn{1}{|c}{\multirow{3}[2]{*}{\textbf{Alg. \newline{}Update}}} & MSE   &             3.9  & \textit{\textbf{            3.8 }} &             4.2  &             6.6  & \textit{\textbf{            3.8 }} &             8.7  &             6.2  &             4.4  &             8.8  &             4.5  \\
          & MAPE  & \textit{\textbf{            3.6 }} &             4.1  &             4.1  &             6.6  &             3.9  &             8.8  &             6.1  &             4.5  &             8.7  &             4.5  \\
          & MAE   & \textit{\textbf{            3.6 }} &             4.1  &             4.2  &             6.6  &             3.9  &             8.8  &             6.1  &             4.4  &             8.9  &             4.5  \\
    \hline
    \multicolumn{1}{|c}{\multirow{3}[2]{*}{\textbf{Meeting}}} & MSE   & \textit{\textbf{            2.9 }} &             4.5  &             3.9  &             7.0  &             4.1  &             8.9  &             6.4  &             3.8  &             8.9  &             4.6  \\
          & MAPE  & \textit{\textbf{            2.8 }} &             4.3  &             4.0  &             7.0  &             4.2  &             8.9  &             6.4  &             4.1  &             8.8  &             4.5  \\
          & MAE   & \textit{\textbf{            2.9 }} &             4.3  &             4.0  &             7.2  &             4.2  &             8.9  &             6.4  &             3.8  &             8.8  &             4.5  \\
    \hline
    \multicolumn{1}{|c}{\multirow{3}[2]{*}{\textbf{Overall}}} & MSE   & \textit{\textbf{            3.2 }} &             4.4  &             4.2  &             6.9  &             3.7  &             9.0  &             6.3  &             4.1  &             8.9  &             4.4  \\
          & MAPE  & \textit{\textbf{            2.8 }} &             4.1  &             3.9  &             7.0  &             3.9  &             9.1  &             6.5  &             4.2  &             8.9  &             4.5  \\
          & MAE   & \textit{\textbf{            2.9 }} &             4.2  &             4.0  &             7.0  &             4.0  &             9.1  &             6.4  &             4.1  &             8.8  &             4.5  \\
    \hline
    \end{tabular}%
  \label{tab:meanRanks3mo}%
\end{table*}%
\end{landscape}

\cref{tab:meanRanks3mo} provides insights into the mean ranks of the various forecasting models across all 110 product categories, structured similarly to \cref{tab:3moErr}. Contrary to \cref{tab:3moErr}, the columns in \cref{tab:2moMeanRanks} aren't standardized by the WSTS column. A rank of 1 indicates the model performed the best for that product category based on the respective error measure, thus lower mean ranks are preferred. 

In general, the observations from \cref{tab:2moMeanRanks} are similar to those those from \cref{tab:2MoRelErr}. The forecasts provided by WSTS demonstrate excellence across most error metrics and scenarios. However, there is an exception with WSTS' Algorithmic Update forecasts, where SARIMA and GPR achieved slightly lower average ranks (3.8 vs. 3.9 for WSTS). Considering that the average MSE for the forecasts based on the SARIMA model was 66\% lower than the MSE of WSTS' algorithmic updates, 
it suggests SARIMA's strong performance is primarily driven by specific product categories where it outperforms WSTS
(\cref{sec:Results_additional} contains a detailed discussion on individual product categories). Although several data-driven models exhibited a superior average error in terms of MSE and MAE for the overall forecasts (for example, SARIMA and GPR) this superiority doesn't necessarily translate to lower average ranks, highlighting WSTS' robust performance across product categories.

Comparing only the data-driven forecasts, GPR emerges as the most accurate data-driven model in the Algorithmic Update and Overall scenarios, while Simple Exponential Smoothing (SES) demonstrates the best performance among data-driven models during quarters where WSTS provided expert forecasts (Meeting), particularly in terms of MSE and MAE. ETS achieved a slightly lower average rank in terms of Mean Absolute Percentage Error (MAPE).

One plausible explanation for the strong performance of the industry experts is their access insider information.
In contrast, the data-driven models rely solely on aggregated historical sales data from the specific product category being forecasted. 
Another factor to consider is the timing of the WSTS meetings where expert forecasts are consolidated. These meetings typically occur in the middle of the forecasted quarter, implying that experts are likely aware of their first-month figures.
Moreover, these figures are available simultaneously with the official forecast release. In contrast, the bi-quarterly algorithmic updates do not integrate this information, though they could potentially benefit from it. Consequently, data-driven models that incorporate this timely information, necessitating forecasts with a horizon of only $h = 2$ months, are evaluated in the subsequent subsection.

\subsection{Forecast Performance with Additional Information}
\label{sec:res2Mo}

Considering the timing of the WSTS meetings in the middle of the quarter,  it's reasonable to assume that industry experts take into account the numbers and internal information pertaining to the first month when formulating their forecasts for the first quarter. Moreover, the consolidated results of the first month are released simultaneously with the forecasts by WSTS. Therefore, utilizing all available information for forecasts seems appropriate. This approach allows for an additional month of data to forecast the quarterly result, rendering a forecasting horizon of $h = 2$ months sufficient. 
Considering that one out of three months' numbers don't require estimation, a plausible anticipation would be to observe approximately a 33\% reduction in MSE (assuming an unbiased estimator and uncorrelated errors). Such a decrease could already position several data-driven methods as competitive alternatives to WSTS, which is examined in this section.

% Table generated by Excel2LaTeX from sheet 'Sheet2'
\begin{landscape}
\begin{table*}[htbp]
  \centering
  \caption{Average performance of the data-driven forecasts with horizon $h = 2$ months across all 110 product categories and relative to WSTS. Each row refers to a different error measure, sorted by WSTS' forecast type: algorithmic update, meeting (expert forecast), and overall. Lower values are preferable. The best value per row is bold and italic.}
    \begin{tabular}{|cccccccccccc|}
    \multicolumn{1}{r}{} &       & \textbf{WSTS} & \textbf {SARIMA} & \textbf{ETS} & \textbf{ET} & \textbf{GPR} & \textbf{KNN} & \textbf{RF} & \textbf{SES} & \textbf{SVM} & \multicolumn{1}{c}{\textbf{Ensemble}} \\
    \hline
    \multicolumn{1}{|c}{\multirow{3}[2]{*}{\textbf{Alg. \newline{}Update}}} & MSE   &              1.00  & \textit{\textbf{            0.10 }} &              0.18  &              0.52  & \textit{\textbf{            0.10 }} &              1.31  &                 0.49  &              0.46  &              1.29  &          0.32  \\
          & MAPE  &              1.00  &              0.62  &              0.62  &              0.76  & \textit{\textbf{            0.61 }} &              1.09  &                 0.75  &              0.68  &              1.08  &          0.65  \\
          & MAE   &              1.00  & \textit{\textbf{            0.42 }} &              0.51  &              0.81  &              0.43  &              1.31  &                 0.79  &              0.77  &              1.24  &          0.63  \\
    \hline
    \multirow{3}[2]{*}{\textbf{Meeting}} & MSE   &              1.00  &              0.68  &              0.66  &              1.41  & \textit{\textbf{            0.64 }} &              3.65  &                 1.25  &              0.96  &              3.18  &          0.91  \\
          & MAPE  &              1.00  & \textit{\textbf{            0.61 }} &              0.62  &              0.82  &              0.65  &              1.15  &                 0.79  &              0.65  &              1.15  &          0.68  \\
          & MAE   &              1.00  & \textit{\textbf{            0.72 }} &              0.73  &              1.05  &              0.75  &              1.54  &                 1.00  &              0.88  &              1.46  &          0.85  \\
    \hline
    \multirow{3}[2]{*}{\textbf{Overall}} & MSE   &              1.00  &              0.28  &              0.32  &              0.80  & \textit{\textbf{            0.27 }} &              2.03  &                 0.72  &              0.61  &              1.86  &          0.50  \\
          & MAPE  &              1.00  & \textit{\textbf{            0.61 }} &              0.62  &              0.79  &              0.63  &              1.12  &                 0.77  &              0.66  &              1.12  &          0.66  \\
          & MAE   &              1.00  & \textit{\textbf{            0.56 }} &              0.61  &              0.92  &              0.58  &              1.41  &                 0.89  &              0.82  &              1.34  &          0.73  \\
    \hline
    \end{tabular}%
  \label{tab:2MoRelErr}%
\end{table*}%
\end{landscape}

The average performance of these forecasts relative to WSTS' is presented in \cref{tab:2MoRelErr}, structured equivalently to Table \cref{tab:3moErr}.
\cref{tab:2MoRelErr} presents the average performance of data-driven forecasts with a horizon of $h = 2$ months across 110 product categories, relative to the forecasts provided by WSTS. Each row represents a different error measure, categorized by WSTS' forecast types: algorithmic update, meeting (expert forecast), and overall. As before, lower values indicate better performance, with the best value per row highlighted in bold and italic.

Contrasting these results with those presented in previous table (\cref{tab:3moErr}), several notable differences emerge. 
Firstly, upon a cursory glance of the results, it becomes evident that the data-driven approaches have exhibited markedly superior performance in this context. 
Whereas WSTS' expert forecasts (Meeting) previously outperformed the data-driven forecasts in terms of MSE, MAPE, and MAE, the tables have now turned, with the data-driven forecasts consistently showcasing superior forecast accuracy in the new scenario (with the exception of SVM and KNN for all as well as ET and RF for the Meeting MSE and MAE comparisons).
Specifically, the GPR model achieved a 36\% lower MSE than WSTS' expert forecasts, followed by ETS (34\% lower) and SARIMA (32\% lower).
In terms of MAPE, SARIMA outperformed WSTS' experts by 39\%, followed by ETS (38\% lower), and GPR and SES (both 35\% lower).
Additionally, SARIMA demonstrated the best performance in terms of MAE (28\% lower than WSTS), trailed by ETS (27\% lower) and GPR (25\% lower). Even the simple ensemble, which even incorporates the forecasts of of the worse performing models, surpassed WSTS' experts by 9\% in MSE, 32\% in MAPE, and 15\% in MAE.
This suggests that data-driven models incorporating the latest available information are highly effective in forecasting outcomes within a shorter horizon.
Furthermore, WSTS' expert  forecasts attained superior average performance in terms of MAPE across all three forecast types with a forecasting horizon of $h=3$ months. However, with a reduced horizon of $h=2$ months, the top-performing data-driven forecasts now outperform WSTS by up to 39\%. %this is true for all forecast types
ARIMA and ETS emerged as the top performers, closely followed by GPR.

Secondly, concerning the algorithmic updates, according to \cref{tab:2MoRelErr}, SARIMA and GPR once again emerge as one of the top-performing models. %Concurrently, the performance of GPR has improved. 
In MSE, both SARIMA and GPR achieved errors 90\% lower than WSTS. Additionally, in terms of MAE, SARIMA attained a 58\% lower error, closely followed by GPR with a 57\% reduction. While in terms of MAPE, where WSTS previously outperformed data-driven models, GPR achieved a 39\% lower error, with SARIMA and ETS achieving a 38\% lower MAPE.

%ranks table
\cref{tab:2moMeanRanks} offers insights into the mean ranks of the various forecasting models across all 110 product categories, organized similarly to \cref{tab:3moErr} and \cref{tab:2MoRelErr}. It's important to note that, in contrast to \cref{tab:3moErr} and \cref{tab:2MoRelErr}, the columns in \cref{tab:2moMeanRanks} aren't standardized by the WSTS column. A rank of 1 indicates the model performed the best for that product category based on the respective error measure, thus lower mean ranks are preferred. Overall, the observations from \cref{tab:2moMeanRanks} parallel those from \cref{tab:2MoRelErr}. Models such as SARIMA, ETS, and GPR consistently garnered high ranks. Notably, most data-driven methods outperformed WSTS' expert forecasts, except for KNN and SVM, which exhibited poorer performance. 

\begin{landscape}
\begin{table*}[htbp]
  \centering
  \caption{Mean ranks of the forecasts with horizon h = 2 months across all 110 product categories. Each row refers to a different error measure, sorted by WSTS' forecast type: algorithmic update, meeting (expert forecast), and overall. Lower values are preferable. The best value per row is bold and italic.}
    \begin{tabular}{|cccccccccccc|}
    \multicolumn{1}{r}{} &       & \textbf{WSTS} & \textbf {SARIMA} & \textbf{ETS} & \textbf{ET} & \textbf{GPR} & \textbf{KNN} & \textbf{RF} & \textbf{SES} & \textbf{SVM} & \multicolumn{1}{c}{\textbf{Ensemble}} \\
    \hline
    
    \multicolumn{1}{|c}{\multirow{3}[2]{*}{\textbf{Alg. \newline{}Update}}} & MSE   &             8.0  &             3.6  &             3.7  &             5.8  & \textit{\textbf{            3.3 }} &             8.4  &             5.7  &             3.9  &             8.7  &             3.7  \\
          & MAPE  &             8.2  & \textit{\textbf{            3.4 }} &             3.7  &             5.8  &             3.6  &             8.6  &             5.6  &             4.1  &             8.4  &             3.5  \\
          & MAE   &             8.0  & \textit{\textbf{            3.4 }} &             3.8  &             5.9  &             3.5  &             8.6  &             5.7  &             4.1  &             8.5  &             3.6  \\
    \hline
    \multicolumn{1}{|c}{\multirow{3}[2]{*}{\textbf{Meeting}}} & MSE   &             6.9  &             3.5  & \textit{\textbf{            3.3 }} &             6.5  &             3.7  &             8.9  &             5.9  &             3.5  &             8.5  &             4.2  \\
          & MAPE  &             7.2  & \textit{\textbf{            2.9 }} &             3.4  &             6.7  &             3.4  &             8.8  &             6.2  &             3.7  &             8.6  &             4.2  \\
          & MAE   &             7.0  & \textit{\textbf{            2.9 }} &             3.5  &             6.8  &             3.5  &             8.8  &             6.2  &             3.7  &             8.5  &             4.1  \\
    \hline
    \multicolumn{1}{|c}{\multirow{3}[2]{*}{\textbf{Overall}}} & MSE   &             7.8  & \textit{\textbf{            3.3 }} &             3.4  &             6.3  & \textit{\textbf{            3.3 }} &             8.8  &             5.9  &             3.6  &             8.7  &             3.8  \\
          & MAPE  &             8.0  & \textit{\textbf{            2.9 }} &             3.2  &             6.4  &             3.3  &             9.0  &             5.9  &             3.8  &             8.7  &             3.7  \\
          & MAE   &             7.8  & \textit{\textbf{            2.9 }} &             3.3  &             6.4  &             3.4  &             9.0  &             6.0  &             3.8  &             8.6  &             3.8  \\
    \hline
    \end{tabular}%
  \label{tab:2moMeanRanks}%
\end{table*}%
\end{landscape}

Finally, \cref{fig:bestModel_all} in the Appendix illustrates the frequency of the best-performing forecasts across 110 WSTS product categories, with colors indicating performance metrics: red for Mean Squared Error (MSE), blue for Mean Absolute Percentage Error (MAPE), and green for Mean Absolute Error (MAE). The left panel reflect the $h=3$ months forecast while $h=2$ is presented on the right side. It's evident that WSTS forecasts rarely emerge as the top performers within any given product category for $h=2$. While WSTS' expert forecasts generally outperform its algorithmic updates, data-driven models consistently outshine both. When comparing against WSTS' expert forecasts (second row), SARIMA emerges as the best performer between 20\% (MSE) and 30\% (MAE) of the time, followed by ETS between 19\% (MSE) and 27\% (MAPE). Additionally, GPR and SES each excel between 13\% (MAE and MAPE) and 17\% and 18\% (MSE), respectively. In contrast, WSTS' expert forecasts demonstrate the best performance between 7\% (MAE and MAPE) and 13\% (MSE) of the time.

These findings corroborate the trends observed in the preceding analyses of this section, suggesting the reliability and effectiveness of certain data-driven models over expert forecasts in near-term forecasting scenarios when additional available information is incorporated, supporting the second research hypothesis (H2).

\subsection{Impact of Time Series Length on Forecast Performance}
\label{sec:resTSLength}

\begin{table*}[htbp]
  \centering
  \caption{Overview of the time series lengths by category.}
    \begin{tabular}{|l|rrrr|}
    \multicolumn{1}{c}{} & \multicolumn{1}{c}{\textbf{Count}} & \multicolumn{1}{c}{\textbf{Av. Length}} & \multicolumn{1}{c}{\textbf{Min. Length}} & \multicolumn{1}{c}{\textbf{Max. Length}} \\
    \hline
    \textbf{Long} & 50    & 392   & 392   & 392 \\
    \textbf{Medium} & 31    & 324   & 296   & 359 \\
    \textbf{Short} & 29    & 222   & 92    & 284 \\
    \hline
    \textbf{All} & 110   & 328   & 92    & 392 \\
    \hline
    \end{tabular}%
  \label{tab:TS_lengths}%
\end{table*}%

This section aims to investigate the possibility that expert forecasts perform better on shorter time series due to the limited data available for training data-driven models (Hypothesis (H3)). To explore this third hypothesis, the 110 product categories are divided into long, medium, and short categories based on the length of available observations. An overview of this categorization is presented in Table \ref{tab:TS_lengths}.

In total, the time series had an average length of 328 monthly observations. Among these, the 50 product categories with available monthly data points for the entire examined time period were classified as ``long" time series. The ``medium" category comprised 31 time series with observations ranging from 296 to 359, averaging 324 available data points. The remaining 29 product categories were designated as ``short" time series, with observations ranging from 92 to 284, averaging 222 available data points.

Similar to \cref{tab:2MoRelErr} and \cref{tab:2MoRelErr}, \cref{tab:tsLengthError2Mo} 
provides an overview of the performance metrics Mean Squared Error (MSE), Mean Absolute Percentage Error (MAPE), and Mean Absolute Error (MAE) for the 2-month forecasts. These metrics are segmented based on time series lengths: Short, Medium, and Long, as delineated in \cref{tab:TS_lengths}, and further categorized by WSTS' forecast types: algorithmic update, meeting (expert forecast), and overall.

\begin{landscape}
\begin{table*}[htbp]
\vskip -1 in
  \centering
  \caption{Average performance of the data-driven forecasts with horizon $h = 2$ months across the different time series lengths and relative to WSTS. Each row refers to a different error measure, sorted by WSTS' forecast type: algorithmic update, meeting (expert forecast), and overall. Lower values are preferable. The best value per row is bold and italic.}
    \begin{tabular}{|c|cccccccccccc|}
    \multicolumn{1}{r}{} &       &       & \textbf{WSTS} & \textbf {SARIMA} & \textbf{ETS} & \textbf{ET} & \textbf{GPR} & \textbf{KNN} & \textbf{RF} & \textbf{SES} & \textbf{SVM} & \multicolumn{1}{c}{\textbf{Ensemble}} \\
    \hline
    \multicolumn{1}{|c|}{\multirow{9}[6]{*}{\begin{sideways}\textbf{Short}\end{sideways}}} & \multicolumn{1}{c}{\multirow{3}[2]{*}{\textbf{Alg. \newline{}Update}}} & MSE   &           1.00  &           0.12  &           0.11  &           0.25  & \textit{\textbf{         0.07 }} &           0.58  &           0.22  &           0.29  &           0.60  &           0.14  \\
          &       & MAPE  &           1.00  &           0.63  &           0.61  &           0.73  & \textit{\textbf{         0.57 }} &           1.08  &           0.72  &           0.63  &           1.07  &           0.63  \\
          &       & MAE   &           1.00  &           0.49  &           0.49  &           0.70  & \textit{\textbf{         0.42 }} &           1.09  &           0.68  &           0.67  &           1.11  &           0.56  \\
\cline{2-13}          & \multirow{3}[2]{*}{\textbf{Meeting}} & MSE   &           1.00  &           0.74  & \textit{\textbf{         0.57 }} &           0.86  & \textit{\textbf{         0.57 }} &           1.35  &           0.77  &           0.65  &           1.26  &           0.66  \\
          &       & MAPE  &           1.00  &           0.69  &           0.66  &           0.88  &           0.77  &           1.23  &           0.85  & \textit{\textbf{         0.65 }} &           1.24  &           0.74  \\
          &       & MAE   &           1.00  &           0.77  & \textit{\textbf{         0.68 }} &           0.89  &           0.74  &           1.23  &           0.85  &           0.72  &           1.20  &           0.76  \\
\cline{2-13}          & \multirow{3}[2]{*}{\textbf{Overall}} & MSE   &           1.00  &           0.33  &           0.27  &           0.46  & \textit{\textbf{         0.24 }} &           0.85  &           0.41  &           0.42  &           0.83  &           0.32  \\
          &       & MAPE  &           1.00  &           0.66  & \textit{\textbf{         0.63 }} &           0.80  &           0.67  &           1.15  &           0.78  &           0.64  &           1.16  &           0.68  \\
          &       & MAE   &           1.00  &           0.63  &           0.58  &           0.79  & \textit{\textbf{         0.57 }} &           1.16  &           0.76  &           0.70  &           1.16  &           0.65  \\
    \hline
    \multicolumn{1}{|c|}{\multirow{9}[6]{*}{\begin{sideways}\textbf{Medium}\end{sideways}}} & \multicolumn{1}{c}{\multirow{3}[2]{*}{\textbf{Alg. \newline{}Update}}} & MSE   &           1.00  & \textit{\textbf{         0.14 }} &           0.17  &           0.42  & \textit{\textbf{         0.14 }} &           1.33  &           0.39  &           0.27  &           1.23  &           0.27  \\
          &       & MAPE  &           1.00  & \textit{\textbf{         0.53 }} &           0.57  &           0.76  &           0.54  &           1.08  &           0.75  &           0.61  &           1.06  &           0.63  \\
          &       & MAE   &           1.00  & \textit{\textbf{         0.45 }} &           0.48  &           0.81  & \textit{\textbf{         0.45 }} &           1.33  &           0.79  &           0.63  &           1.32  &           0.64  \\
\cline{2-13}          & \multirow{3}[2]{*}{\textbf{Meeting}} & MSE   &           1.00  & \textit{\textbf{         0.58 }} &           0.61  &           1.05  &           0.66  &           2.17  &           0.93  &           0.73  &           2.21  &           0.78  \\
          &       & MAPE  &           1.00  & \textit{\textbf{         0.57 }} &           0.65  &           0.83  &           0.64  &           1.20  &           0.79  &           0.66  &           1.15  &           0.67  \\
          &       & MAE   &           1.00  & \textit{\textbf{         0.64 }} &           0.67  &           0.97  &           0.72  &           1.45  &           0.92  &           0.76  &           1.48  &           0.81  \\
\cline{2-13}          & \multirow{3}[2]{*}{\textbf{Overall}} & MSE   &           1.00  & \textit{\textbf{         0.31 }} &           0.34  &           0.66  &           0.34  &           1.66  &           0.60  &           0.45  &           1.61  &           0.47  \\
          &       & MAPE  &           1.00  & \textit{\textbf{         0.55 }} &           0.61  &           0.79  &           0.59  &           1.14  &           0.77  &           0.63  &           1.10  &           0.65  \\
          &       & MAE   &           1.00  & \textit{\textbf{         0.54 }} &           0.57  &           0.89  &           0.58  &           1.39  &           0.85  &           0.69  &           1.39  &           0.72  \\
    \hline
    \multicolumn{1}{|c|}{\multirow{9}[6]{*}{\begin{sideways}\textbf{Long}\end{sideways}}} & \multicolumn{1}{c}{\multirow{3}[2]{*}{\textbf{Alg. \newline{}Update}}} & MSE   &           1.00  & \textit{\textbf{         0.10 }} &           0.18  &           0.54  & \textit{\textbf{         0.10 }} &           1.34  &           0.50  &           0.47  &           1.31  &           0.32  \\
          &       & MAPE  &           1.00  & \textit{\textbf{         0.67 }} &           0.68  &           0.78  &           0.68  &           1.09  &           0.77  &           0.76  &           1.11  &           0.68  \\
          &       & MAE   &           1.00  & \textit{\textbf{         0.41 }} &           0.52  &           0.83  &           0.43  &           1.34  &           0.81  &           0.80  &           1.25  &           0.64  \\
\cline{2-13}          & \multirow{3}[2]{*}{\textbf{Meeting}} & MSE   &           1.00  &           0.68  &           0.66  &           1.45  & \textit{\textbf{         0.64 }} &           3.80  &           1.28  &           0.98  &           3.29  &           0.93  \\
          &       & MAPE  &           1.00  &           0.58  &           0.58  &           0.78  & \textit{\textbf{         0.57 }} &           1.07  &           0.75  &           0.64  &           1.09  &           0.64  \\
          &       & MAE   &           1.00  & \textit{\textbf{         0.72 }} &           0.75  &           1.08  &           0.76  &           1.60  &           1.03  &           0.93  &           1.50  &           0.88  \\
\cline{2-13}          & \multirow{3}[2]{*}{\textbf{Overall}} & MSE   &           1.00  & \textit{\textbf{         0.27 }} &           0.33  &           0.81  & \textit{\textbf{         0.27 }} &           2.08  &           0.74  &           0.63  &           1.91  &           0.51  \\
          &       & MAPE  &           1.00  & \textit{\textbf{         0.62 }} &           0.63  &           0.78  & \textit{\textbf{         0.62 }} &           1.08  &           0.76  &           0.70  &           1.10  &           0.66  \\
          &       & MAE   &           1.00  & \textit{\textbf{         0.55 }} &           0.63  &           0.94  &           0.58  &           1.46  &           0.91  &           0.86  &           1.36  &           0.75  \\\hline
    \end{tabular}%
  \label{tab:tsLengthError2Mo}%
\end{table*}%
\end{landscape}

For short time series, GPR demonstrated superior performance in terms of both MSE and MAPE, showcasing reductions of 76\% and 43\% respectively compared to WSTS' combined forecasts. Regarding MAPE, ETS exhibited the lowest error rate (37\% lower than WSTS), trailed by SES (36\% lower), SARIMA (34\% lower), and GPR (33\% lower). When contrasting the data-driven forecasts with WSTS' expert forecasts, ETS emerged as the top-performing model, tying with GPR in terms of MSE (both 43\% lower than expert forecasts) and outperforming WSTS' experts by 32\% in terms of MAE. However, SES marginally outperformed ETS in terms of MAPE, showing reductions of 35\% and 34\% respectively. When comparing data-driven forecasts solely to algorithmic forecasts, GPR emerged as the most accurate model across all metrics, boasting reductions of 93\% in MSE, 43\% in MAPE, and 58\% in MAE. This explains the robust performance of GPR relative to WSTS' combined forecasts.

It is also noteworthy to highlight the disparity in the performance of data-driven methods for short time series depending on whether quarters with expert forecasts or algorithmic updates were used for the benchmarks. Given that these are derived from the same time series, a consistent ranking of data-driven methods might have been expected. The best-performing models for medium and long time series remain largely consistent, regardless of whether they are evaluated using algorithmic forecasts or expert forecasts. Even in cases where there are differences, the margin is minimal.

In the Appendix, \cref{tab:tsLengthRanks} outlines the mean ranks of the analyzed models, once again categorized by Mean Squared Error (MSE), Mean Average Percentage Error (MAPE), and Mean Absolute Error (MAE), and segmented by time series lengths. These rankings are further delineated by WSTS forecast type, mirroring the structure of \cref{tab:tsLengthError2Mo}.

Consistent with the findings in \cref{tab:tsLengthError2Mo}, GPR demonstrated the most favorable performance in terms of mean ranks across MSE (2.8), MAPE (2.8), and MAE (2.9) for short time series when compared to WSTS' algorithmic forecasts. However, while ETS excelled in average MSE and MAE, and SES in MAPE, this outcome is reversed when considering mean ranks: SES exhibited the best performance in terms of mean ranks for MSE and MAE, while ETS performed best for MAPE. It's worth noting that the differences between them are minimal in each case. Furthermore, ETS attained the lowest mean rank when all quarters were taken into account (overall), suggesting that exponential smoothing models perform well when data is limited. Furthermore, the high accuracy of the data-driven forecasts compared to WSTS' expert forecasts is evidence contrary to the third research hypothesis (H3) that expert forecasts would outperform data-driven forecast in the context of short time series.     

In fact, a slight increase in error relative to the expert forecasts is observed for the best forecasts when medium-length time series are considered in terms of MSE (0.58 vs. 0.57). Likewise, the data-driven forecasts fared slightly worse in terms of MSE (0.64 vs. 0.57) and MAE (0.72 vs. 0.68) for long time series. Given the additional information which was utilized in training the models, the opposite might have been expected. This was the case when the forecasts were evaluated by MAPE (0.57 medium and long time series vs. 0.65 for short ones). 

For medium time series, SARIMA generated the most reliable forecasts in terms of relative average error measures -- 42\% more accurate than the experts polled by WSTS in terms of MSE, 43\% more accurate in terms of MAPE, and 36\% more accurate in terms of MAE, followed by ETS. This is also reflected in the mean ranks of the SARIMA forecasts (2.9, 2.2, and 2.8 vs. 7.1, 7.8, and 7.5 respectively). Similarly, the SARIMA forecasts demonstrated superior performance in comparison to the algorithmic updates and when evaluated overall.

In the case of the long time series, the most accurate forecasts resulted from the GPR and SARIMA models (\cref{tab:tsLengthError2Mo}). GPR resulted in 36\% lower MSE and a 43\% lower MAPE compared to the experts. SARIMA excelled in terms of MAE -- 38\% lower than WSTS. A similar picture arises when mean ranks (\cref{tab:tsLengthRanks}) are considered. An exception is the algorithmic update category, where forecasts based on the ensemble method achieved the lowest mean ranks.

\cref{fig:bestModel_TS} in the Appendix illustrates the frequency of the best-performing forecasts with a forecast horizon of $h = 2$ months. The chart is divided into 3 columns corresponding to the time series lengths: short, medium, and long. Rows are arranged by WSTS forecast type (algorithmic update, meeting (expert), and overall). The colors differentiate between Mean Squared Error (MSE) in green, Mean Absolute Percentage Error (MAPE) in blue, and Mean Absolute Error (MAE) in red. In concordance with the analysis of \cref{tab:tsLengthError2Mo} and \cref{tab:tsLengthRanks}, GPR shows the highest frequency of top model performance for short sime series when compared to the algorithmic updates. Overall and for the expert forecasts ETS and SES had the highest frequencies of highest accuracy. For the medium and long time series, SARIMA, GPR, and the exponential smoothing models excelled most often. 

Additional details pertaining to the overall performance of the various data-driven models for each product category categorized by time series lengths is available in \cref{tab:errByProd}.

\subsection{Additional Results}
\label{sec:Results_additional}

In addition the the aggregated results presented in \cref{sec:resQuart} - \cref{sec:resTSLength}, this Section elaborates the results on a product category level. 

\cref{tab:rmseByProd}, \cref{tab:errByProd}, and \cref{tab:maeByProd} in the Appendix, provide the Root Mean Squared Errors (RMSE), which is the square root of the Mean Squared Error (MSE) described in \cref{sec:Methodology} and was chosen for readability here, mean absolute percentage errors (MAPE), and Mean Absolute Errors (MAE) respectively for various forecasting models across 110 different product categories.
Each time, these are organized by time series lengths, consistent with the categorization in \cref{sec:resTSLength}. The error measures for the forecasts with horizon $h = 3$ months are presented in the first half of each table and those forecasts with horizon $h = 2$ months are presented in the second halfs.
Lower RMSE, MAPE, and MAE values indicate higher forecasting accuracy, with the best performing model per product category printed in bold and italic in each row of each table.

In the 3-month forecast category, the results are consistent with those discussed in \cref{sec:resQuart}. As can be seen in \cref{tab:rmseByProd} and \cref{tab:maeByProd}, data-driven methods, particularly GPR and SARIMA, are overall able to outperform WSTS in terms of RMSE and MAE. \cref{tab:rmseByProd} reveals that this is in large part due to the strong performace of the GPR ans SARIMA forecasts in terms of RMSE for a few product categories such as M99 (37.72 for GPR vs 46.77 for WSTS), T99 (56.83 for GPR vs. 71.85 for WSTS), and S2 (49.17 for SARIMA vs 70.79 for WSTS). Similarly, scrutinizing \cref{tab:maeByProd} indicates that among all forecasts, WSTS' was the most reliable for most product categories in terms of MAE. But GPR, SARIMA, and ETS produced forecasts which excelled for specific product categories, such as T99 (53.76 for WSTS vs. 46.24 for SARIMA, 47.35 for GPR, and 50.85 for ETS), resulting in a higher average performance for long time series. In terms of MAPE, \cref{tab:errByProd} illustrates WSTS' strong performance  on the $h = 3$ month horizon across all time series lengths. Nevertheless, upon scrutinizing \cref{tab:errByProd}, it becomes apparent that data-driven models outperform WSTS' combined expert and algorithmic update forecasts in terms of MAPE for specific product categories. For instance, in categories such as J99 and L8a, the SARIMA model produces the lowest MAPE, demonstrating its effectiveness in forecasting these particular products.
In some categories like L1, where the time series might exhibit unique patterns or complexities, traditional statistical models such as SARIMA and ETS perform inadequately compared to specific machine learning models, in this case: RF.

Consistent with the observations in \cref{sec:res2Mo}, these results are reversed when the forecasts with a horizon of $h = 2$ months are considered. In this scenario, data-driven forecasts excelled across the vast majority of product categories in terms of RMSE (\cref{tab:rmseByProd}), MAPE (\cref{tab:errByProd}), and MAE (\cref{tab:maeByProd}). Nevertheless, WSTS' forecasts were superior for some select product categories with long histories in terms of RMSE, such as C7 (1.18 vs. 1.32 for the best performing data-driven model: GPR), CC (1.29 vs. 1.47 for the best performing data-driven model: GPR), and S3 (2.77 vs. 2.94 for the best performing data-driven model: SES).

Moreover, despite the dominance of the SARIMA, GPR, and ETS among the data-driven models, it is noteworthy that other models with poorer average performance still yielded in strong forecasts for select product categories: ET was amongst the top performing models for the A5 product category in terms of RMSE and MAPE, and SVM performed best for the JCd product category in terms of RMSE and MAE. 
This finding is underlined by \cref{fig:bestModel_all} located in the Appendix. The bar chart visualizes the frequency of the best-performing forecasts across 110 WSTS product categories. Differentiated by colors, green signifies the best performance based on Mean Squared Error (MSE), blue represents Mean Absolute Percentage Error (MAPE), and red indicates Mean Absolute Error (MAE). The chart is divided into two sections: the left side displays outcomes for forecasts with a horizon of $h = 3$ months, while the right side portrays forecasts with a horizon of $h = 2$ months (to be discussed in the subsequent section). Rows are arranged by algorithmic update, expert forecasts (meeting), and overall performance. 
%Analyzing the frequency at which WSTS' forecasts outperformed the data-driven models underscores the reliability of the official predictions in this setting, particularly those derived from industry experts.

Finally, \cref{fig:bestModel_all} in the Appendix illustrates the frequency of the best-performing forecasts across 110 WSTS product categories, with colors indicating performance metrics: red for Mean Squared Error (MSE), blue for Mean Absolute Percentage Error (MAPE), and green for Mean Absolute Error (MAE). The left panel reflect the $h=3$ months forecast while $h=2$ is presented on the right side. 

For $h=3$ months the superior performance of the WSTS models is evident, particularly their Meeting and Overall forecasts. Depending on the error measure, WSTS' experts outperformed all data driven models between 46\% and 48\% of all product categories. The most frequent best performing data driven models were SARIMA, GPR and the exponential smoothing models, ETS and SES. In aggregate, the forecasts provided by WSTS (Overall row) achieved a top performance in 38\% to 39\% of the cases. In contrast, the playing field was more even when only the algorithmic updates are considered. WSTS achieved a top performance for 21\% to 23\% of all product categories, followed by SARIMA with 17\% to 23\% and GPR with 13\% to 19\% of all product categories.

It's evident that WSTS forecasts rarely emerge as the top performers within any given product category for $h=2$ months. While WSTS' expert forecasts generally outperform its algorithmic updates, data-driven models consistently outshine both. When comparing against WSTS' expert forecasts (second row), SARIMA emerges as the best performer between 20\% (MSE) and 30\% (MAE) of the time, followed by ETS between 19\% (MSE) and 27\% (MAPE). Additionally, GPR and SES each excel between 13\% (MAE and MAPE) and 17\% and 18\% (MSE), respectively. In contrast, WSTS' expert forecasts demonstrate the best performance between 7\% (MAE and MAPE) and 13\% (MSE) of the time.

This highlights that there is no one model for all situations but that the model choice should depend on the individual product category and the error measure with the greatest business relevance.

\section{Discussion}
\label{sec:discussion}

\cref{sec:Introduction} made the case that the semiconductor industry plays a crucial role in the broader economy and stressed the importance of reliable forecasts for operational and strategic decision making. Furthermore, the rapidly evolving technologies, complicated geopolitical considerations, and complex supply chains exposing the industry to the bullwhip effect make data-driven forecasting more challenging. Concurrently, industry insiders, such as those queried by the World Semiconductor Trade Statistics (WSTS) -- a leading provider of semiconductor market data and forecasts, promise reliable forecasts based on a wealth of quantitative and qualitative insider information.

This motivated the first research hypothesis (H1) that expert forecasts exhibit higher accuracy compared to data-driven forecasts. This hypothesis was extensively examined in \cref{sec:resQuart} for a forecast horizon of $h=3$. The analysis of the benchmark concluded that the bi-quarterly expert forecasts indeed demonstrated superior accuracy on a quarterly horizon in terms of Mean Squared Error (MSE), Mean Absolute Percentage Error (MAPE), and Mean Absolute Error (MAE). In contrast to the superior performance of the expert forecasts, it was also found that the bi-quarterly algorithmic forecasts provided by WSTS showed potential for further improvement. 

Furthermore, it was noted that industry insiders may have access to additional information owing to the timing of WSTS meetings, during which the forecasts are formulated. This observation prompted the formulation of the second research hypothesis (H2) that the additional information yields competitive data-driven forecasts, which was investigated in \cref{sec:res2Mo}. It was found that the additional data and the shorter horizon of $h = 2$ months significantly improved forecasts based on quantitative models. Forecasts based on the SARIMA, GPR, ETS, and SES models consistently demonstrated superior accuracy. As a consequence, it is recommended that practitioners complement expert forecasts with data-driven methods to enhance the forecasts reliability.

General wisdom holds that experts excel in situations with limited historical data \cite{Hyndman.2018Book}. %hyndman book - cited in introduction
Consequently, the third research hypothesis (H3), which postulates that industry insiders outperform in short time series due to restricted quantitative data available for model training, was examined in \cref{sec:resTSLength}. The analysis revealed that data-driven forecasts exhibited superiority across all examined time series lengths. Nonetheless, different models showcased higher accuracy under varying circumstances. Specifically, exponential smoothing models attained the highest accuracy for short time series, whereas SARIMA dominated in the medium-length scenario. Conversely, GPR outperformed for longer time series. This implies that several diverse models should be evaluated and the one that aligns most effectively with the given circumstances should be selected.

These findings demonstrate the potential of data-driven forecasts, even in a dynamic environment with numerous potential extrinsic influences like the semiconductor industry. The forecasts sourced from the WSTS provide a strong baseline, which can be improved upon with quantitative modelling. Moreover, data-driven forecasts can easily be updated with new information, leading to faster incorporation of current data and trends. It is worth mentioning that integrating insider information into data-driven models holds promise for enhancing accuracy and tailoring forecasts to the unique context of each specific company. The central position of the semiconductor industry in the wider economy and the need for reliable forecasts for operational and strategic planning underscore the importance of these findings.

\appendix
\section{Mean Ranks by Time Series Lengths}

\begin{landscape}
\begin{table*}[htbp]
\vskip -1 in
  \centering
  \caption{Mean ranks of the forecasts with horizon h = 2 months across the different time series lengths. Each row refers to a different error measure, sorted by WSTS' forecast type: algorithmic update, meeting (expert forecast), and overall. Lower values are preferable. The best value per row is bold and italic.}
\begin{tabular}{|c|cccccccccccc|}
    \multicolumn{1}{r}{} &       &       & \textbf{WSTS} & \textbf {SARIMA} & \textbf{ETS} & \textbf{ET} & \textbf{GPR} & \textbf{KNN} & \textbf{RF} & \textbf{SES} & \textbf{SVM} & \multicolumn{1}{c}{\textbf{Ensemble}} \\
    \hline
    \multicolumn{1}{|c|}{\multirow{9}[6]{*}{\begin{sideways}\textbf{Short}\end{sideways}}} & \multicolumn{1}{c}{\multirow{3}[2]{*}{\textbf{Alg. \newline{}Update}}} & MSE   &             8.1  &             3.4  &             3.3  &             5.9  & \textit{\textbf{            2.8 }} &             8.9  &             5.9  &             4.0  &             8.9  &             3.9  \\
          &       & MAPE  &             8.2  &             3.6  &             3.5  &             6.1  & \textit{\textbf{            2.8 }} &             8.8  &             5.8  &             4.0  &             8.5  &             3.7  \\
          &       & MAE   &             7.9  &             3.3  &             3.4  &             6.1  & \textit{\textbf{            2.9 }} &             8.9  &             5.9  &             4.1  &             8.7  &             3.9  \\
\cline{2-13}          & \multicolumn{1}{c}{\multirow{3}[2]{*}{\textbf{Meeting}}} & MSE   &             7.2  &             3.8  &             2.8  &             6.6  &             4.0  &             9.1  &             5.9  & \textit{\textbf{            2.7 }} &             8.6  &             4.2  \\
          &       & MAPE  &             7.2  &             3.3  & \textit{\textbf{            2.8 }} &             6.8  &             3.8  &             8.9  &             6.3  &             3.0  &             8.7  &             4.3  \\
          &       & MAE   &             7.0  &             3.3  &             3.0  &             6.7  &             4.1  &             8.9  &             6.1  & \textit{\textbf{            2.9 }} &             8.8  &             4.2  \\
\hline
    \end{tabular}%
  \label{tab:tsLengthRanks}%
\end{table*}%
\end{landscape}

\begin{landscape}
\section{Error Measures by Product Category}
\subsection{RMSE}

\begingroup
  \centering
    \begin{longtable}{l|rrrrrrrrrr}
      \caption{Root Mean Sqared error (RMSE), the square root of the Mean Squared Error (MSE), for the 110 examined WSTS product categories, which are organized by time series length and presented for the 3-month forecasts (first) and 2-month forecast (second). Averages are contained in the long, medium, short, 3-months, and 2-months rows respectively. The best result in each row, i.e. for each time series, is printed bold and italic. The RMSE was chosen instead of the MSE for readability purposes in this table.}
          \label{tab:rmseByProd}%
      \\
      
    \multicolumn{1}{c}{\textbf{ ID }} & \multicolumn{1}{c}{\textbf{WSTS}} & \multicolumn{1}{c}{\textbf {SARIMA}} & \multicolumn{1}{c}{\textbf{ETS}} & \multicolumn{1}{c}{\textbf{ET}} & \multicolumn{1}{c}{\textbf{GPR}} & \multicolumn{1}{c}{\textbf{KNN}} & \multicolumn{1}{c}{\textbf{RF}} & \multicolumn{1}{c}{\textbf{SES}} & \multicolumn{1}{c}{\textbf{SVM}} & \multicolumn{1}{c}{\textbf{Ensemble}} \\
        \hline
\endfirsthead
    \hline
    \multicolumn{1}{c}{\textbf{ ID }} & \multicolumn{1}{c}{\textbf{WSTS}} & \multicolumn{1}{c}{\textbf {SARIMA}} & \multicolumn{1}{c}{\textbf{ETS}} & \multicolumn{1}{c}{\textbf{ET}} & \multicolumn{1}{c}{\textbf{GPR}} & \multicolumn{1}{c}{\textbf{KNN}} & \multicolumn{1}{c}{\textbf{RF}} & \multicolumn{1}{c}{\textbf{SES}} & \multicolumn{1}{c}{\textbf{SVM}} & \multicolumn{1}{c}{\textbf{Ensemble}} \\
        \hline
\endhead
    \textbf{3 Months} &                   4.14  &                   3.87  &                   4.07  &                   5.51  & \textit{\textbf{                  3.78 }} &                   8.93  &                   5.36  &                   4.50  &                   8.61  &                   4.27  \\
    \hline
    \hline
    \textbf{long} &                   6.87  &                   6.24  &                   6.71  &                   9.39  & \textit{\textbf{                  6.06 }} &                 15.55  &                   9.11  &                   7.59  &                 14.76  &                   7.08  \\
    \hline
    \hline
    A2    &                   0.17  &                   0.19  & \textit{\textbf{                  0.15 }} &                   0.25  &                   0.17  &                   0.38  &                   0.23  &                   0.16  &                   0.51  &                   0.19  \\
    A5    & \textit{\textbf{                  0.11 }} &                   0.18  &                   0.18  &                   0.13  &                   0.15  &                   0.22  &                   0.13  &                   0.15  &                   0.20  &                   0.15  \\
    A99   & \textit{\textbf{                  0.48 }} &                   0.56  &                   0.53  &                   0.80  & \textit{\textbf{                  0.48 }} &                   1.29  &                   0.82  &                   0.53  &                   1.16  &                   0.59  \\
    B2    & \textit{\textbf{                  0.18 }} &                   0.20  &                   0.21  &                   0.28  &                   0.24  &                   0.45  &                   0.27  &                   0.20  &                   0.53  &                   0.19  \\
\end{longtable}%
\endgroup

\subsection{MAPE}

\begingroup
  \centering
    \begin{longtable}{l|rrrrrrrrrr}
      \caption{Mean absolute percentage errors (MAPE) for the 110 examined WSTS product categories, which are organized by time series length and presented for the 3-month forecasts (first) and 2-month forecast (second). Averages are contained in the long, medium, short, 3-months, and 2-months rows respectively. The best result in each row, i.e. for each time series, is printed bold and italic.}
          \label{tab:errByProd}%
      \\
      
    \multicolumn{1}{c}{\textbf{ ID }} & \multicolumn{1}{c}{\textbf{WSTS}} & \multicolumn{1}{c}{\textbf {SARIMA}} & \multicolumn{1}{c}{\textbf{ETS}} & \multicolumn{1}{c}{\textbf{ET}} & \multicolumn{1}{c}{\textbf{GPR}} & \multicolumn{1}{c}{\textbf{KNN}} & \multicolumn{1}{c}{\textbf{RF}} & \multicolumn{1}{c}{\textbf{SES}} & \multicolumn{1}{c}{\textbf{SVM}} & \multicolumn{1}{c}{\textbf{Ensemble}} \\
        \hline
\endfirsthead
    \hline
    \multicolumn{1}{c}{\textbf{ ID }} & \multicolumn{1}{c}{\textbf{WSTS}} & \multicolumn{1}{c}{\textbf {SARIMA}} & \multicolumn{1}{c}{\textbf{ETS}} & \multicolumn{1}{c}{\textbf{ET}} & \multicolumn{1}{c}{\textbf{GPR}} & \multicolumn{1}{c}{\textbf{KNN}} & \multicolumn{1}{c}{\textbf{RF}} & \multicolumn{1}{c}{\textbf{SES}} & \multicolumn{1}{c}{\textbf{SVM}} & \multicolumn{1}{c}{\textbf{Ensemble}} \\
        \hline
\endhead
    \textbf{3 Months} & \textit{\textbf{8.4\%}} & 9.0\% & 8.8\% & 10.6\% & 9.0\% & 14.5\% & 10.3\% & 9.0\% & 14.7\% & 9.1\% \\
    \hline
    \hline
    \textbf{long} & \textit{\textbf{7.7\%}} & 8.1\% & 8.1\% & 9.6\% & 8.0\% & 12.7\% & 9.5\% & 8.5\% & 13.3\% & 8.3\% \\
    \hline
    \hline
    A2    & \textit{\textbf{5.5\%}} & 6.6\% & 5.9\% & 8.3\% & 7.3\% & 12.7\% & 7.7\% & 6.1\% & 15.5\% & 6.8\% \\
    A5    & \textit{\textbf{4.6\%}} & 7.1\% & 6.5\% & 5.4\% & 6.3\% & 9.1\% & 5.7\% & 6.5\% & 8.3\% & 5.9\% \\
    A99   & \textit{\textbf{4.2\%}} & 5.2\% & 5.5\% & 7.6\% & 5.0\% & 12.3\% & 7.5\% & 5.4\% & 10.3\% & 5.8\% \\
    B2    & 5.2\% & \textit{\textbf{4.9\%}} & 5.2\% & 9.1\% & 6.1\% & 13.4\% & 8.8\% & 5.7\% & 16.3\% & 6.4\% \\
    B4    & \textit{\textbf{6.0\%}} & 8.5\% & 7.4\% & 7.6\% & 7.5\% & 9.7\% & 7.8\% & 6.8\% & 7.8\% & 7.3\% \\
    B99   & \textit{\textbf{4.6\%}} & 6.5\% & 5.8\% & 5.2\% & 5.6\% & 5.7\% & 5.5\% & 5.8\% & 5.3\% & 5.3\% \\
\end{longtable}%
\endgroup

\subsection{MAE}

\begingroup
  \centering
    \begin{longtable}{l|rrrrrrrrrr}
      \caption{Mean absolute errors (MAE) for the 110 examined WSTS product categories, which are organized by time series length and presented for the 3-month forecasts (first) and 2-month forecast (second). Averages are contained in the long, medium, short, 3-months, and 2-months rows respectively. The best result in each row, i.e. for each time series, is printed bold and italic.}
          \label{tab:maeByProd}%
      \\
      
    \multicolumn{1}{c}{\textbf{ ID }} & \multicolumn{1}{c}{\textbf{WSTS}} & \multicolumn{1}{c}{\textbf {SARIMA}} & \multicolumn{1}{c}{\textbf{ETS}} & \multicolumn{1}{c}{\textbf{ET}} & \multicolumn{1}{c}{\textbf{GPR}} & \multicolumn{1}{c}{\textbf{KNN}} & \multicolumn{1}{c}{\textbf{RF}} & \multicolumn{1}{c}{\textbf{SES}} & \multicolumn{1}{c}{\textbf{SVM}} & \multicolumn{1}{c}{\textbf{Ensemble}} \\
        \hline
\endfirsthead
    \hline
    \multicolumn{1}{c}{\textbf{ ID }} & \multicolumn{1}{c}{\textbf{WSTS}} & \multicolumn{1}{c}{\textbf {SARIMA}} & \multicolumn{1}{c}{\textbf{ETS}} & \multicolumn{1}{c}{\textbf{ET}} & \multicolumn{1}{c}{\textbf{GPR}} & \multicolumn{1}{c}{\textbf{KNN}} & \multicolumn{1}{c}{\textbf{RF}} & \multicolumn{1}{c}{\textbf{SES}} & \multicolumn{1}{c}{\textbf{SVM}} & \multicolumn{1}{c}{\textbf{Ensemble}} \\
        \hline
\endhead
    \textbf{3 Months} &                3.12  &           3.01  &           3.10  &           4.43  & \textit{\textbf{         2.97 }} &           6.93  &           4.27  &           3.59  &           6.59  &           3.44  \\
    \hline
    \hline
    \textbf{long} &                5.24  &           4.92  &           5.18  &           7.63  & \textit{\textbf{         4.83 }} &         12.03  &           7.35  &           6.18  &         11.27  &           5.82  \\
    \hline
    \hline
    A2    & \textit{\textbf{              0.11 }} &           0.13  & \textit{\textbf{         0.11 }} &           0.16  &           0.14  &           0.26  &           0.15  &           0.12  &           0.34  &           0.13  \\
    A5    & \textit{\textbf{              0.09 }} &           0.14  &           0.13  &           0.10  &           0.12  &           0.18  &           0.11  &           0.12  &           0.16  &           0.11  \\
    A99   & \textit{\textbf{              0.34 }} &           0.43  &           0.44  &           0.62  &           0.40  &           1.03  &           0.62  &           0.43  &           0.90  &           0.48  \\
    B2    &                0.14  & \textit{\textbf{         0.13 }} &           0.14  &           0.24  &           0.17  &           0.36  &           0.23  &           0.15  &           0.44  &           0.17  \\
    B4    & \textit{\textbf{              0.13 }} &           0.20  &           0.17  &           0.16  &           0.16  &           0.21  &           0.17  &           0.15  &           0.17  &           0.16  \\
    B99   & \textit{\textbf{              0.25 }} &           0.36  &           0.33  &           0.28  &           0.32  &           0.32  &           0.30  &           0.31  &           0.28  &           0.29  \\
\end{longtable}%
\endgroup

\end{landscape}

\section{Best Model Frequencies}

\begin{figure}[ht]
\vskip 0.2in
\begin{center}
\centerline{\includegraphics[width=\columnwidth]{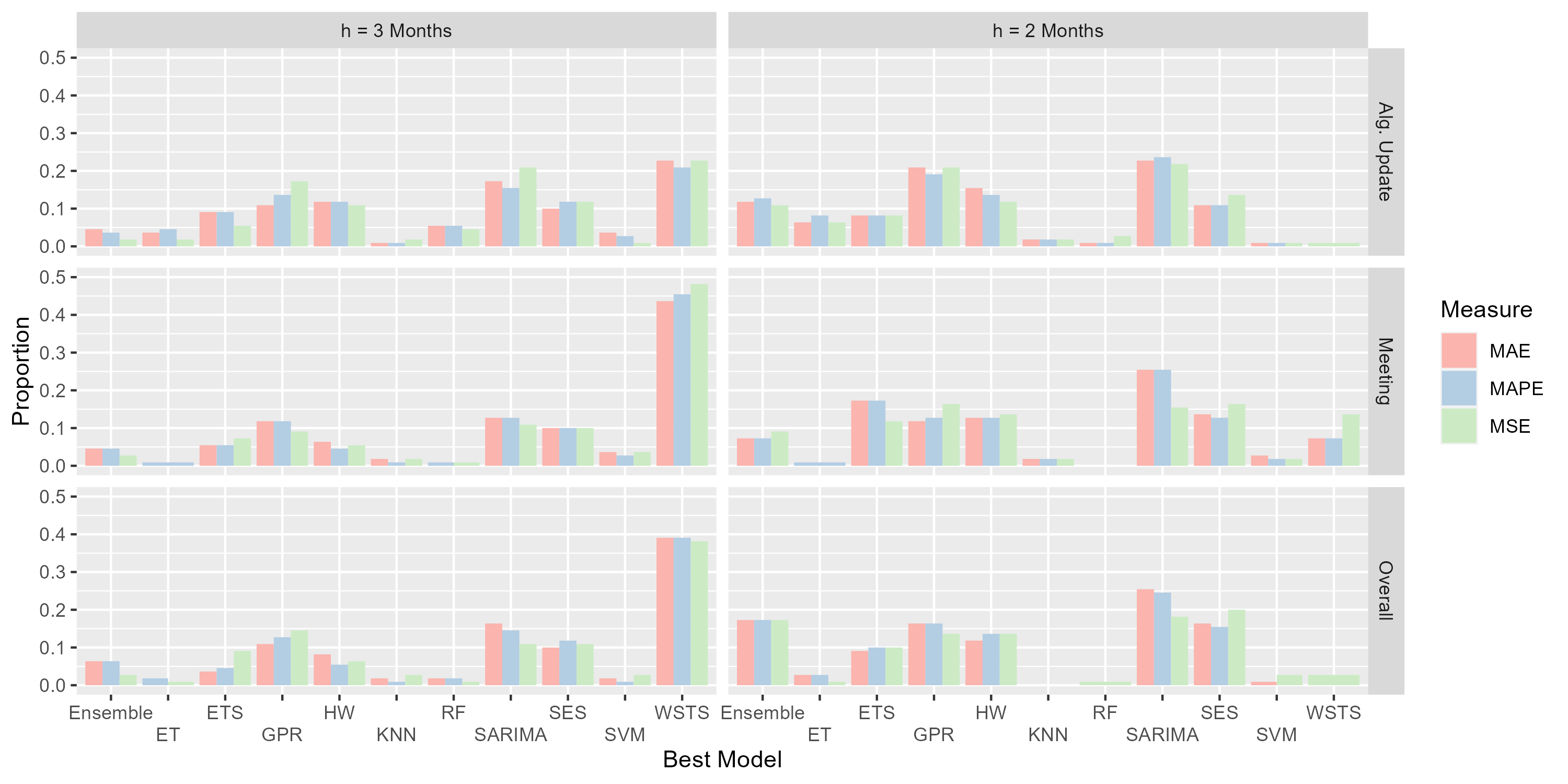}}
\caption{Bar chart illustrating the frequency of the best-performing forecasts based on Mean Squared Error (MSE) in red, Mean Absolute Percentage Error (MAPE) in blue, and Mean Absolute Error (MAE) in green, for the 110 WSTS product categories. The left side of the chart represents the outcomes of forecasts with a horizon of $h = 3$ months, while the right side represents those with a forecast horizon of $h = 2$ months. The rows are organized by algorithmic update, expert forecasts ("Meeting"), and overall performance, mirroring the structure of \cref{tab:3moErr} and \cref{tab:2MoRelErr}.
}
\label{fig:bestModel_all}
\end{center}
\vskip -0.2in
\end{figure}

\begin{figure}[ht]
\vskip 0.2in
\begin{center}
\centerline{\includegraphics[width=\columnwidth]{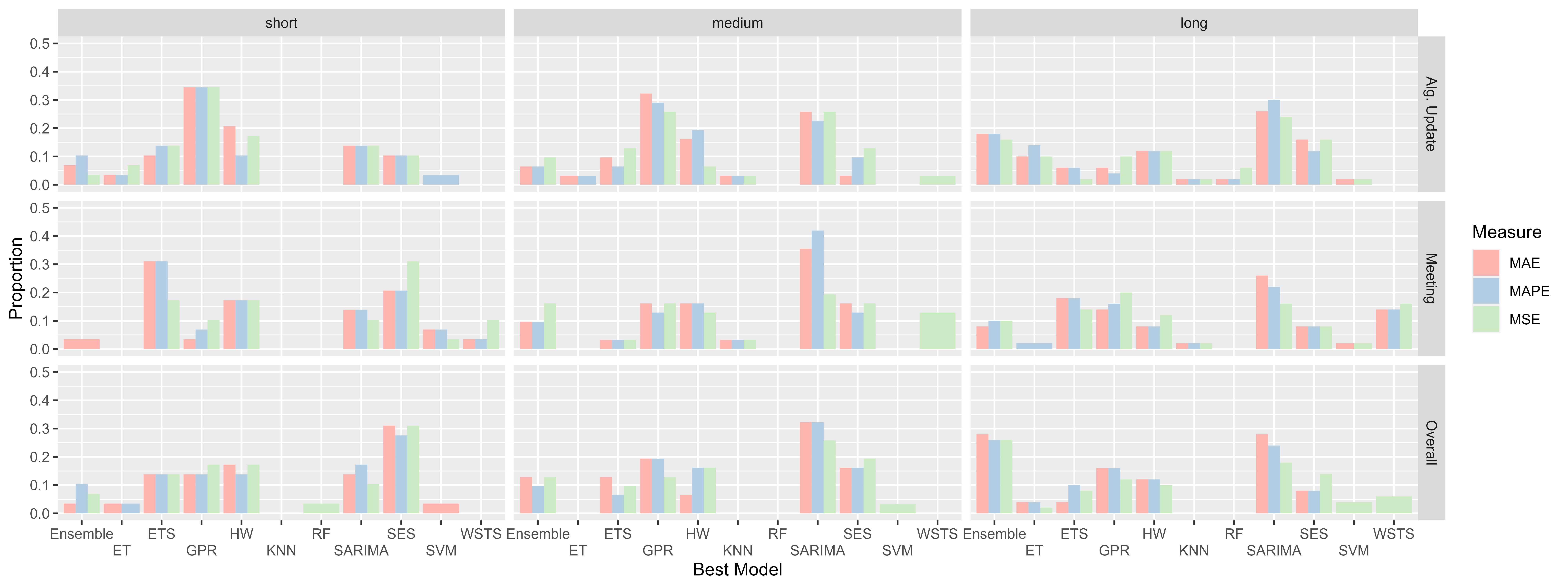}}
\caption{The bar chart provides insight into the comparative performance of forecasting methods across different time series lengths and evaluation metrics.
It depicts the distribution of the best-performing forecasts across 110 WSTS product categories, categorized by Mean Squared Error (MSE) shown in red, Mean Absolute Percentage Error (MAPE) in blue, and Mean Absolute Error (MAE) in green. Each column represents a different time series length category: the left column represents short time series, the middle column represents medium-length series, and the rightmost column represents long series (those with full available history).
The rows of the chart are arranged by algorithmic update, expert forecasts ("Meeting"), and overall performance, following the structure of \cref{tab:3moErr} and \cref{tab:2MoRelErr}.
}
\label{fig:bestModel_TS}
\end{center}
\vskip -0.2in
\end{figure}

\section*{Acknowledgements}
This paper was supported by a research cooperation between Infineon Technologies AG and TU Dortmund University through the Graduate School of Logistics.

\section*{Declaration of generative AI and AI-assisted technologies in the writing process}

During the preparation of this work the authors used DALL$\cdot$E 3 in order to generate an appealing graphical abstract. Furthermore, the ChatGPT 3.5 model from OpenAI was employed for minor language edits, aiming to enhance readability. After using this tool/service, the authors reviewed and edited the content as needed and take full responsibility for the content of the publication.

\pagebreak
\onecolumn



%%       \citet{<label>} ==> Jones et al. [21]
%%       \citep{<label>} ==> [21]
%%

 \bibliographystyle{elsarticle-num-names} 
 \bibliography{HumanVsMachine.bib}





% \end{thebibliography}
\end{document}